\documentclass[fleqn,usenatbib]{mnras}

\usepackage{newtxtext,newtxmath}
\usepackage[T1]{fontenc}
\usepackage{ae,aecompl}

\usepackage{arydshln}
\usepackage{bm}
\usepackage{comment}
\usepackage{multirow}

\usepackage{graphicx}	% Including figure files
\usepackage{amsmath}	% Advanced maths commands
\usepackage{amssymb}	% Extra maths symbols

\title[Kernel phase imaging with VLT/NACO]{Kernel phase imaging with VLT/NACO: high-contrast detection of new candidate low-mass stellar companions at the diffraction limit}

\author[J. Kammerer et al.]{
Jens Kammerer,$^{1}$\thanks{E-mail: jens.kammerer@anu.edu.au}
Michael J. Ireland,$^{1}$
Frantz Martinache$^{2}$
and Julien H. Girard$^{3,4}$
\\
$^{1}$Research School of Astronomy \& Astrophysics, Australian National University, ACT 2611, Australia\\
$^{2}$Laboratoire Lagrange, Universit\'e C\^ote d'Azur, Observatoire de la C\^ote d'Azur, CNRS, Parc Valrose, B\^at. H. FIZEAU, 06108 Nice, France\\
$^{3}$Space Telescope Science Institute, 3700 San Martin Drive, Baltimore, MD 21218, USA\\
$^{4}$Universit\'e Grenoble Alpes, CNRS, IPAG, 38000 Grenoble, France
}

\date{Accepted XXX. Received YYY; in original form ZZZ}

\pubyear{2018}

\begin{document}
\label{firstpage}
\pagerange{\pageref{firstpage}--\pageref{lastpage}}
\maketitle

\begin{abstract}
Directly imaging exoplanets is challenging because quasi-static phase aberrations in the pupil plane (speckles) can mimic the signal of a companion at small angular separations. Kernel phase, which is a generalization of closure phase (known from sparse aperture masking), is independent of pupil plane phase noise to second order and allows for a robust calibration of full pupil, extreme adaptive optics observations.
We applied kernel phase combined with a principal component based calibration process to a suitable but not optimal, high cadence, pupil stabilized L' band ($3.8~\text{\textmu m}$) data set from the ESO archive. We detect eight low-mass companions, five of which were previously unknown, and two have angular separations of $\sim0.8$--$1.2~\lambda/D$ (i.e. $\sim80$--$110~\text{mas}$), demonstrating that kernel phase achieves a resolution below the classical diffraction limit of a telescope. While we reach a $5\sigma$ contrast limit of $\sim1/100$ at such angular separations, we demonstrate that an optimized observing strategy with more diversity of PSF references (e.g. star-hopping sequences) would have led to a better calibration and even better performance.
As such, kernel phase is a promising technique for achieving the best possible resolution with future space-based telescopes (e.g. JWST), which are limited by the mirror size rather than atmospheric turbulence, and with a dedicated calibration process also for extreme adaptive optics facilities from the ground.
\end{abstract}

\begin{keywords}
planets and satellites: detection -- planets and satellites: formation -- techniques: high angular resolution -- techniques: image processing -- techniques: interferometric -- binaries: close
\end{keywords}

%%%%%%%%%%%%%%%%%%%%%%%%%%%%%%%%%%%%%%%%%%%%%%%%%%

%%%%%%%%%%%%%%%%% BODY OF PAPER %%%%%%%%%%%%%%%%%%

\section{Introduction}
\label{sec:introduction}

Direct imaging is vital for studying the outer regions of extrasolar systems which are inaccessible to transit observations and can only be revealed by decades-long, time consuming radial velocity surveys \citep[e.g.][]{fischer2014}. It has proven particularly successful in probing our understanding of the formation of gas giant planets \citep[e.g.][]{d'angelo2010}, being able to estimate their mass from their luminosity and age \citep[e.g.][]{spiegel2012} and resolve their orbit. Although the majority of detected companion candidates are arguably consistent with being emission or scattering from disk material (e.g. LkCa~15, \citealt{kraus2012}, HD~100546, \citealt{quanz2013}, HD~169142, \citealt{biller2014}), the recent example of PDS~70 \citep{keppler2018} demonstrates that direct imaging of wide-separation but still solar-system scale planets is possible at relatively moderate contrasts in the vicinity of young stars. This is spurring an ongoing discussion about the nature of planet formation and the commonness of gas giant planets with large orbital distances \citep[e.g.][]{bowler2018}.

However, direct imaging operates at the resolution and sensitivity limit of the most powerful instruments today \citep[e.g.][]{pepe2014}, placing demanding requirements on the observing and the post-processing techniques which are used to uncover faint companions at high contrasts (e.g. angular differential imaging, \citealt{marois2006}, point spread function subtraction, \citealt{lafreniere2007a}, principal component analysis, \citealt{amara2012}, \citealt{soummer2012}). Detecting exoplanets from the ground using these techniques has only been made possible by the recent development of extreme adaptive optics systems \citep[e.g.][]{milli2016} and is mainly limited by non-common path aberrations which are not sensed by the wavefront control system \citep[e.g.][]{sauvage2007}. These aberrations manifest themselves as quasi-static speckles on the detector images which can mimic the signal of a companion and place a strong constraint on the achievable contrast at small angular separations \citep[e.g.][]{fitzgerald2006}. Hence, directly imaging and studying the formation of gas giant planets on solar-system scales has been extremely challenging so far \citep[e.g.][]{bowler2016} because the nearest star forming regions lie $\gtrsim100~\text{pc}$ away \citep[e.g.][]{loinard2007} where such orbital distances correspond to angular separations of only $\lesssim200~\text{mas}$.

In this paper, we explore the capabilities of the kernel phase technique \citep{martinache2010} for high-contrast imaging at the diffraction limit from the ground. This post-processing technique can be seen as refinement of sparse aperture masking and the closure phase technique \citep{tuthill2000}. By probing only certain linear combinations of the phase of the Fourier transformed detector images, kernel phase and sparse aperture masking allow for a robust calibration of the time-varying optical transfer function of the system and a significant mitigation of quasi-static speckles and achieve an angular resolution of $\lesssim50~\text{mas}$ in the near-infrared \citep[i.e. the L' band,][]{cheetham2016}. This gives access to objects on solar-system scales in the nearest star forming regions (i.e. projected separations of $\sim40~\text{mas}$ for a Jupiter analog in the Scorpius Centaurus OB association, \citealt{preibisch2008}) and has proven successful in directly imaging young exoplanets/disk features \citep[e.g.][]{kraus2012}. The caveat of sparse aperture masking is that the mask blocks $\gtrsim85~\%$ of the light \citep[for VLT/NACO,][]{tuthill2010} and therefore significantly decreases the sensitivity and hence the contrast limit of the observations for relatively faint targets. However, kernel phase uses the light collected by the entire pupil and should perform better in the high Strehl regime and the bright limit \citep[e.g.][]{pope2016, sallum2019}.

For sparse aperture masking, a mask is placed at the Lyot stop of an instrument in order to split the primary mirror into a discrete interferometric array of \textit{real} sub-apertures \citep[e.g.][]{readhead1988}. In the Fourier transform of the detector image (hereafter referred to as Fourier plane), these sub-apertures map onto their auto-correlation (i.e. their spatial frequencies, \citealt{ireland2016}). The phase $\phi$ of each spatial frequency can be extracted and linearly combined in a way such that the resulting closure phase $\theta = \bm{K}\cdot\phi$ is independent of the pupil plane (or instrumental) phase $\varphi$ to second order (i.e. terms of first and second order in $\varphi$ are vanishing), where the matrix $\bm{K}$ encodes this special linear combination \citep[e.g.][]{ireland2016}. For observations from the ground, the pupil plane phase $\varphi$ is affected by noise from atmospheric seeing and non-common path aberrations which ultimately cause quasi-static speckles. Being more robust with respect to these systematic effects, sparse aperture masking achieves a superior angular resolution.

For full pupil kernel phase imaging, there is no mask and the entire primary mirror is discretized into an interferometric array of \textit{virtual} sub-apertures. According to \citet{martinache2010}, it is then convenient to define a transfer matrix $\bm{A}$ which maps the baselines between each pair of \textit{virtual} sub-apertures onto their corresponding spatial frequency. In the high Strehl regime, where the pupil plane phase $\varphi$ can be linearized, we obtain the relationship
\begin{equation}
    \label{eqn:phase}
    \phi = \bm{R}^{-1}\cdot\bm{A}\cdot\varphi+\phi_\text{obj}+\mathcal{O}(\varphi^3),
\end{equation}
where $\bm{R}$ is a diagonal matrix encoding the redundancy of the spatial frequencies (i.e. the baselines of the interferometric array) and $\phi_\text{obj}$ is the phase intrinsic to the observed object. Multiplication with the left kernel $\bm{K}$ of $\bm{R}^{-1}\cdot\bm{A}$ yields
\begin{align}
    \label{eqn:kernel_phase_begin}
    \theta &= \bm{K}\cdot\phi\\
           &= \underbrace{\bm{K}\cdot\bm{R}^{-1}\cdot\bm{A}}_{= 0}\cdot\varphi+\bm{K}\cdot\phi_\text{obj}+\mathcal{O}(\varphi^3)\\
    \label{eqn:kernel_phase_end}
           &= \theta_\text{obj}+\underbrace{\mathcal{O}(\varphi^3)}_{\approx 0},
\end{align}
hence the kernel $\theta$ of the measured Fourier plane phase $\phi$ directly represents the kernel $\theta_\text{obj}$ of the phase intrinsic to the observed object $\phi_\text{obj}$, at least in the high Strehl regime (where $\mathcal{O}(\varphi^3)$ is negligible). This is why frame selection based on the Strehl ratio is essential. Note that the kernel phase is a generalization of the closure phase to the case of redundant apertures.

For observations from space, which do not suffer from atmospheric seeing, kernel phase has proven to be successful in resolving close companions at the diffraction limit \citep{martinache2010,pope2013}. It is our goal to determine if, under good observing conditions, kernel phase also is a competitive alternative to sparse aperture masking from the ground.

\section{Methods}
\label{sec:methods}

\subsection{Data reduction}
\label{sec:data_reduction}

A basic direct imaging data reduction (such as dark, flat, background subtraction and bad pixel correction) is also essential for the kernel phase technique \citep[e.g.][]{sallum2017}. For this purpose, we developed our own data reduction pipeline\footnote{\url{https://github.com/kammerje/PyConica}} which can be fed the raw data with their associated raw calibrators from the VLT/NACO archive\footnote{\url{http://archive.eso.org/wdb/wdb/eso/naco/form}}. Our data reduction pipeline performs the following steps which are described in more detail in the following sections:
\begin{enumerate}
    \item Linearize the raw frames.
    \item Compute master darks and their bad pixel maps.
    \item Compute master flats and their bad pixel maps.
    \item Flag saturated pixels.
    \item Apply dark, flat, background and bad pixel corrections.
    \item Perform a dither subtraction.
    \item Reconstruct saturated pixels.
    \item Select frames with sufficient Strehl ratio.
\end{enumerate}

\subsubsection{Detector linearization correction}
\label{sec:detector_linearization_correction}

\begin{figure*}
    \centering
    \includegraphics[width=\textwidth]{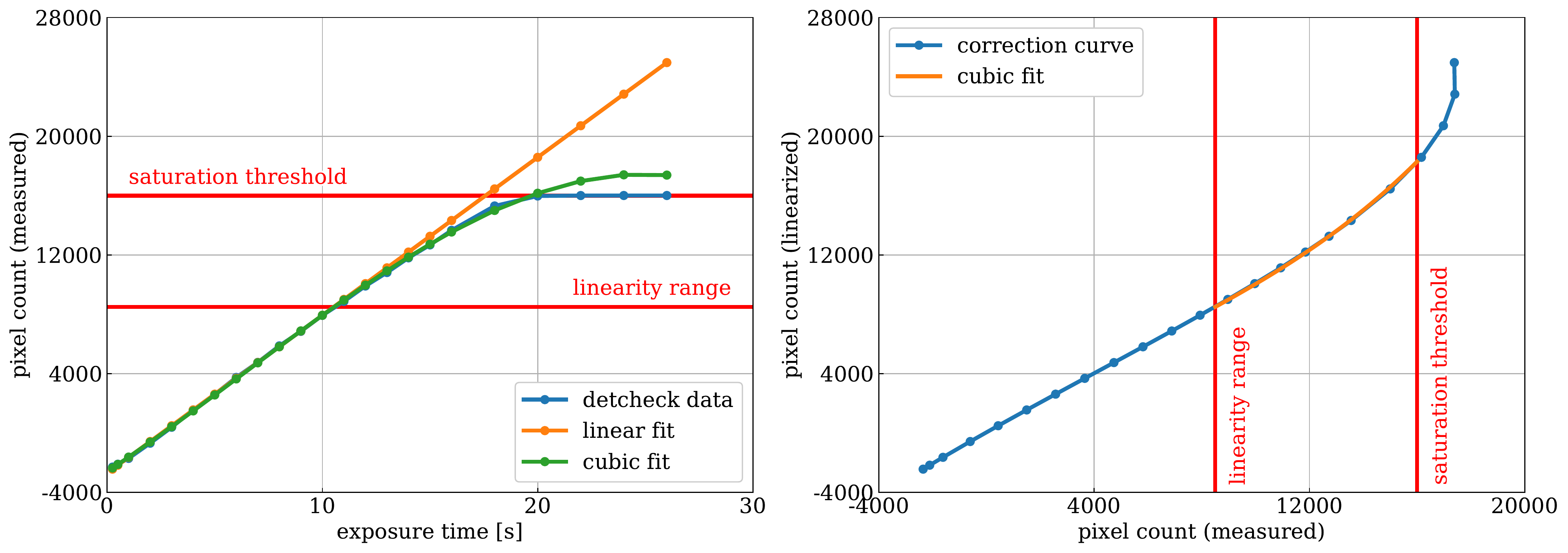}
    \caption{Left panel: median pixel count in dependence of the integration time $t$ for uncorrelated high well depth mode from the detector monitoring (blue curve) and the linear (orange curve) and cubic (green curve) polynomials $f(t)$ and $g(t)$ which we fit to it. Right panel: correction curve (blue) $f(g)$ and the cubic polynomial $h$ (orange curve) which we fit to it and use for linearizing all pixels with measured counts between 8500 and 16000. In both panels, the solid red lines mark the end of the linear regime and the saturation threshold. Note that very low (i.e. negative) pixel counts occur due to the use of a narrow-band filter ($\Delta\lambda = 0.018~\text{\textmu m}$) for the detector monitoring, whereas the L' science frames are taken with a wide-band filter ($\Delta\lambda = 0.62~\text{\textmu m}$).}
    \label{fig:detector_linearization_correction}
\end{figure*}

Like most photon counting devices, NACO's infrared detector CONICA suffers from a non-linear response when the pixel counts approach the saturation threshold (16400 counts for uncorrelated high well depth mode\footnote{This is the standard imaging mode in the L' band ($3.8~\text{\textmu m}$) and all data cubes which we analyze have been taken in this mode.} according to the NACO Quality Control and Data Processing\footnote{\url{https://www.eso.org/observing/dfo/quality/NACO/qc/detmon_qc1.html}}, with a more conservative 16000 counts used in our analysis). As kernel phase is an interferometric technique for which the fringes are coded spatially on the detector, it is very important to characterize the pixel to pixel response. Moreover, many of the data cubes which we analyze in Section~\ref{sec:results_discussion} feature saturated point spread functions (PSFs) which we want to correct for non-linearity before reconstructing their core (cf. Section~\ref{sec:reconstruction_saturated_pixels}).

In order to compute the detector linearization correction we download all frames of type ``FLAT, LAMP, DETCHECK'' and uncorrelated high well depth mode from 2016 March 23 and 2016 September 24 (which are closest in time to the observation of the earliest and the latest data cube which we analyze) from the VLT/NACO archive. We sort them by integration time and compute the median pixel count over all frames for each individual integration time (masking out the broken stripes in the lower left quadrant of CONICA). Then, we plot the median pixel count in dependence of the integration time $t$, fit a linear polynomial $f(t)$ to all data points with less than 8500 counts (end of the linear regime for uncorrelated high well depth mode) and a cubic polynomial $g(t)$ to all data points with less than 16000 counts (saturation threshold, cf. left panel of Figure~\ref{fig:detector_linearization_correction}). We linearize the detector using a continuously differentiable piecewise polynomial approach $h$ to the correction curve $f(g)$ with a linear function up to 8500 counts and a cubic polynomial between 8500 and 16000 counts (cf. right panel of Figure~\ref{fig:detector_linearization_correction}).

\subsubsection{Master darks and master flats}
\label{sec:master_darks_master_flats}

For each observation block (OB) we compute master darks from the associated dark frames as the median of all dark frames with a unique set of size and exposure time. Then we compute a bad pixel map for each master dark based on the frame by frame median and variance of each pixel's count. Therefore, we first compute two frames:
\begin{enumerate}
    \item The absolute difference between the master dark and the median filtered master dark.
    \item The absolute of the median subtracted variance dark.
\end{enumerate}
Then, we identify bad pixels in each of these frames based on their difference to the median of these frames. For frame (i) we classify pixels which are above 10 times the median as bad, for frame (ii) pixels which are above 75 times the median. Note that these thresholds were identified empirically. From the median subtracted dark frames, we estimate the readout noise as the mean over each frame's pixel count standard deviation.

We proceed similar for the flat frames, but also group them by filter as well as size and exposure time, subtract a master dark with matching properties (i.e. similar size and exposure time) from each master flat and normalize it by its median pixel count.

\subsubsection{Saturated pixels}
\label{sec:saturated_pixels}

The data cubes which we analyze in Section~\ref{sec:results_discussion} consist of 100 frames of $0.2~\text{s}$ exposure. For each data cube, we reject the first frame (which we find to consistently suffer from a bias), so that there are 99 frames left. Note that NACO appends the median of all 100 frames at the end of each data cube which is also rejected here. Before proceeding, we also flag the saturated pixels in each frame which are all pixels with more than $h(16000)$ counts.

\subsubsection{Dark, flat, background and bad pixel correction}
\label{sec:dark_flat_background_bad_pixel_correction}

\begin{figure}
    \centering
    \includegraphics[width=\columnwidth]{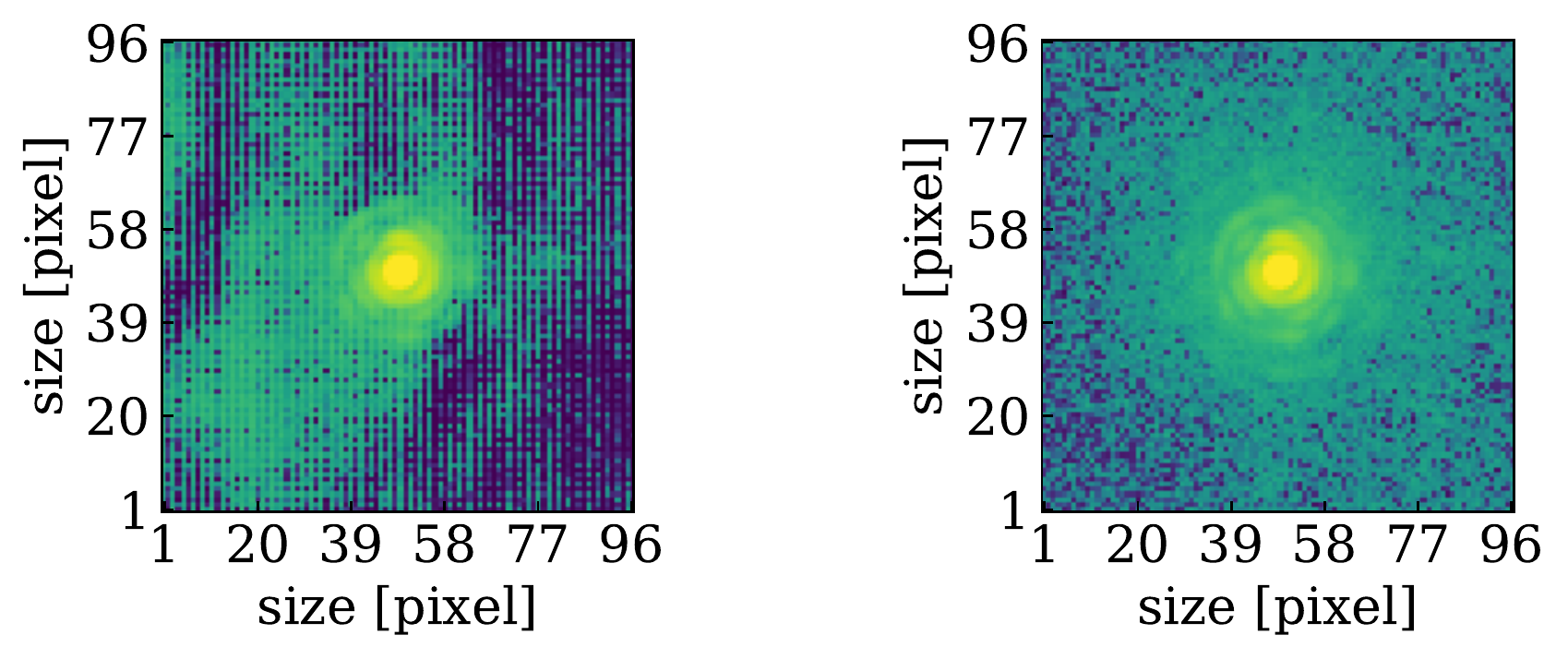}
    \caption{Left panel: median frame of a data cube of HIP~47425 after dark, flat and a simple background subtraction. The pixel counts are scaled by an arcsinh stretch so that both the PSF and the background are visible in the image. Right panel: same median frame after performing the dither subtraction described in Section~\ref{sec:dither_subtraction}. This second step is essential to remove residual systematic noise from the detector which can be seen as grid-like structure in the left panel. Note that the two panels have the same color scale.}
    \label{fig:hip_47425_dark_flat_background_subtraction}
\end{figure}

We clean each frame of a data cube individually by subtracting a master dark with matching properties (i.e. similar size and exposure time), dividing it by a master flat with matching properties (i.e. similar size, exposure time and filter), correcting bad pixels (which are bad pixels from the master dark or the master flat) with a median filter of size five pixels and performing a simple background subtraction by subtracting the median pixel count of the frame from each pixel. A typical result is shown in the left panel of Figure~\ref{fig:hip_47425_dark_flat_background_subtraction}, where residual systematic noise (mainly from the detector) is still clearly visible.

\subsubsection{Dither subtraction}
\label{sec:dither_subtraction}

In order to mitigate the residual systematic noise from the detector and the sky background we perform a dither subtraction. After cleaning all data cubes within one OB, we find for each data cube (which we will here call data cube A) the data cube B with the target furthest away (on the detector) and subtract its median frame from each frame of data cube A. The new bad and saturated pixel maps are then the logical sums of those from both involved data cubes. After this step the residual noise appears like Gaussian random noise as is shown in the right panel of Figure~\ref{fig:hip_47425_dark_flat_background_subtraction}.

Our typical performance is a pixel count standard deviation of $\sim36 = 4.4+(158/\text{s}\cdot0.2~\text{s})$ outside of $10~\lambda/D$ from the center of the PSF in $0.2~\text{s}$ exposure, where $4.4$ is the detector readout noise, $\lambda$ is the observing wavelength ($3.8~\text{\textmu m}$ for the L' band) and $D$ is the diameter of the primary mirror ($8.2~\text{m}$ for the VLT).

\subsubsection{Reconstruction of saturated pixels}
\label{sec:reconstruction_saturated_pixels}

\begin{figure}
    \centering
    \includegraphics[width=\columnwidth]{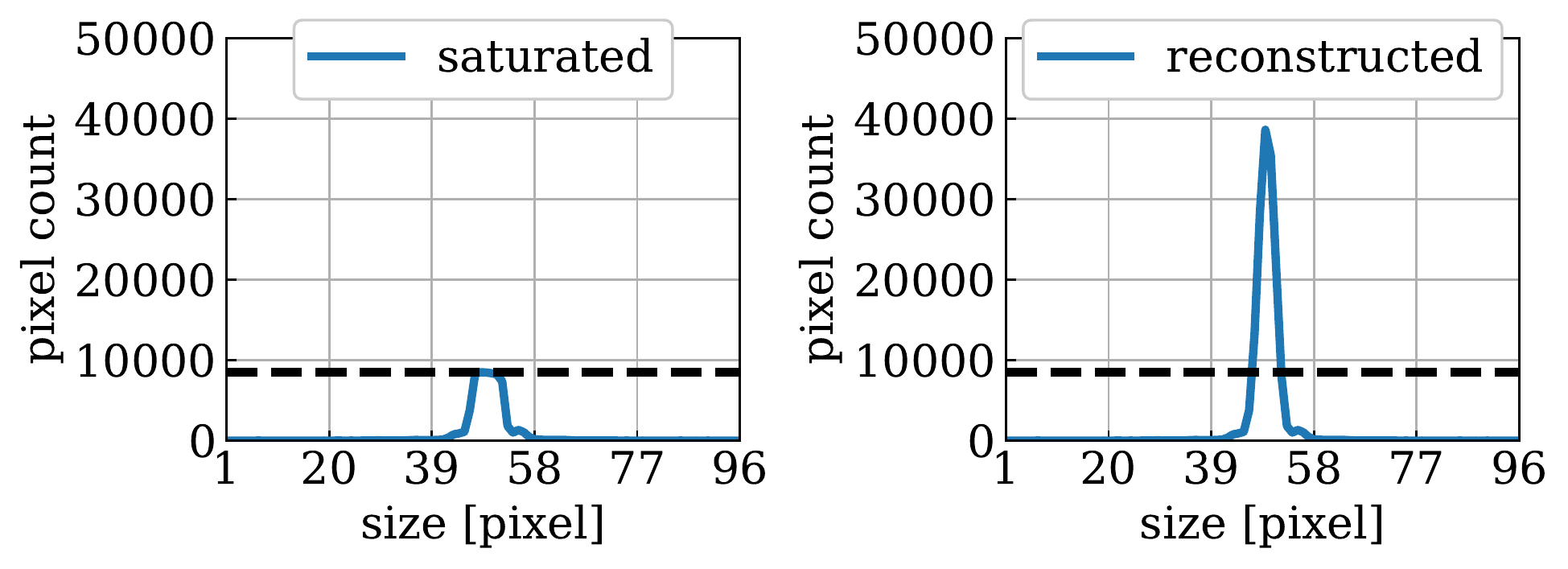}
    \caption{Left panel: mean over a horizontal and a vertical cross-section through the center of the median frame shown in the right panel of Figure~\ref{fig:hip_47425_dark_flat_background_subtraction}. Right panel: same cross-section, but after reconstructing bad and saturated pixels as described in Section~\ref{sec:reconstruction_saturated_pixels}. The dashed black line marks the maximum of the cross-section in the left panel in order to illustrate the reconstruction of the peak in the PSF core.}
    \label{fig:hip_47425_cross_section}
\end{figure}

\begin{figure}
    \centering
    \includegraphics[width=\columnwidth]{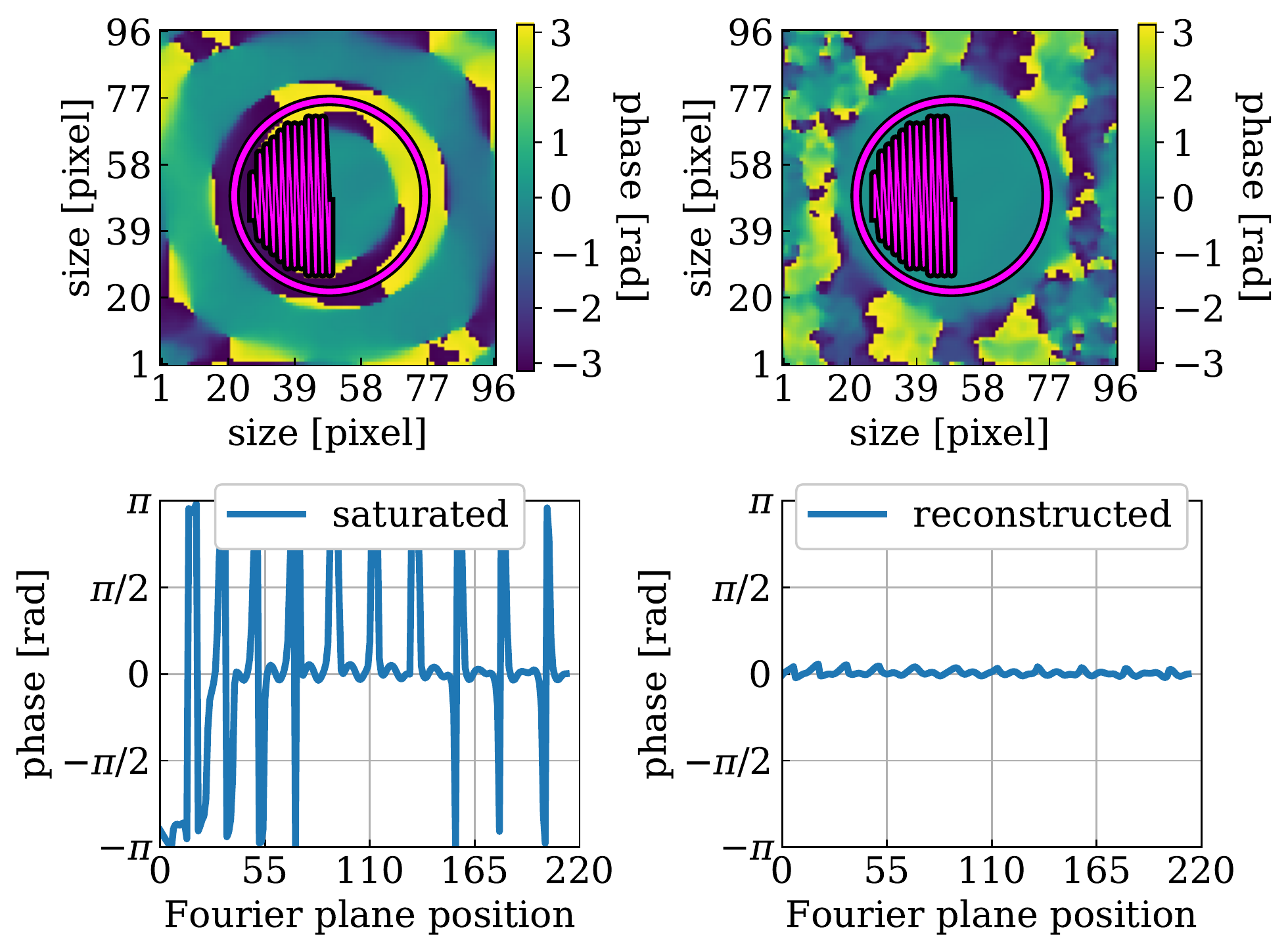}
    \caption{Left panels: Fourier plane phase of the median frame shown in the right panel of Figure~\ref{fig:hip_47425_dark_flat_background_subtraction} (top). The phase is flat in the center, but the cutoff spatial frequency is smaller than the region of support permitted by the pupil geometry (magenta circle). Median Fourier plane phase at the spatial frequencies of our pupil model (bottom). Right panels: same as in the left panels, but after reconstructing bad and saturated pixels as described in Section~\ref{sec:reconstruction_saturated_pixels}. In both upper panels, the magenta line traces out the spatial frequencies of our pupil model (from left to right) in order to illustrate how the patterns observed in the lower panels are obtained.}
    \label{fig:hip_47425_fourier_phase}
\end{figure}

Our reconstruction of saturated pixels is based on an algorithm described in Section~2.5 of \citet{ireland2013}. This technique also identifies and corrects residual bad pixels, with no more than 10 additional bad pixels corrected in a typical frame. First, we crop all frames to a size of 96 by 96 pixels ($\sim2.6~\text{arcsec}^2$) centered on the target. Then, we correct bad and saturated pixels for each frame separately by minimizing the Fourier plane power $\left|f_Z\right|$ outside the region of support $Z$ permitted by the pupil geometry. Let $\bm{B}_Z$ be the matrix which maps the bad and saturated pixel values $x$ onto the Fourier plane domain $Z$, then
\begin{equation}
    f_Z = \bm{B}_Z\cdot{b}+\epsilon_Z,
\end{equation}
where $b$ are the corrections to the bad and saturated pixel values $x$ (i.e. the corrected pixel values are $x-b$) and $\epsilon_Z$ is remaining Fourier plane noise. We solve for $b$ using the Moore-Penrose pseudo-inverse of $\bm{B}_Z$, i.e.
\begin{equation}
    b = \bm{B}_Z^+\cdot{f_Z} = \left(\bm{B}_Z^*\cdot\bm{B}_Z\right)^{-1}\cdot\bm{B}_Z^*\cdot{f_Z}.
\end{equation}
Since a broad-band filter was used for the observations, but we use a monochromatic central filter wavelength in our analysis and also blur the edge of the pupil through the use of a windowing function, we use a sightly larger pupil diameter to define this region $Z$ of $10~\text{m}$ here. In fact, the only important thing for recovering the Fourier plane phase is that the Fourier plane power \textit{outside} the region of support permitted by the pupil geometry is minimized, so using a larger pupil diameter just assures this in case of low quality data and is a conservative choice, especially in the case of our data which is far from the Nyquist sampling criterion.

Sometimes, the remaining Fourier plane noise $\epsilon_Z$ can be significant, which is why we repeat the entire correction process up to 15 times for each frame. After each iteration, we look for remaining bad pixels by:
\begin{enumerate}
    \item Computing the Fourier transform of the corrected frame from the previous iteration.
    \item Windowing this frame by the Fourier domain $Z$.
    \item Computing the inverse Fourier transform of this frame.
    \item Identifying remaining bad pixels in this frame based on their difference to the median filtered frame.
\end{enumerate}
If no remaining bad pixels are identified, we terminate the iteration.

A cross-section of a saturated PSF before and after performing the reconstruction is shown in Figure~\ref{fig:hip_47425_cross_section}. Obviously, this reconstruction cannot reveal any structure or companions hidden behind saturated pixels, but it allows us to perform our kernel phase analysis on saturated data cubes which would otherwise suffer from high Fourier plane phase noise (cf. Figure~\ref{fig:hip_47425_fourier_phase}). Please note that a method from the class of least squares spectral analysis techniques (i.e. image plane fringe fitting) may be more robust in dealing with bad pixels, but would require the simultaneous fitting of all Fourier plane phases and amplitudes and is therefore beyond the scope of this paper, although it is a promising approach for future work.

\subsubsection{Frame selection}
\label{sec:frame_selection}

\begin{figure}
    \centering
    \includegraphics[width=\columnwidth]{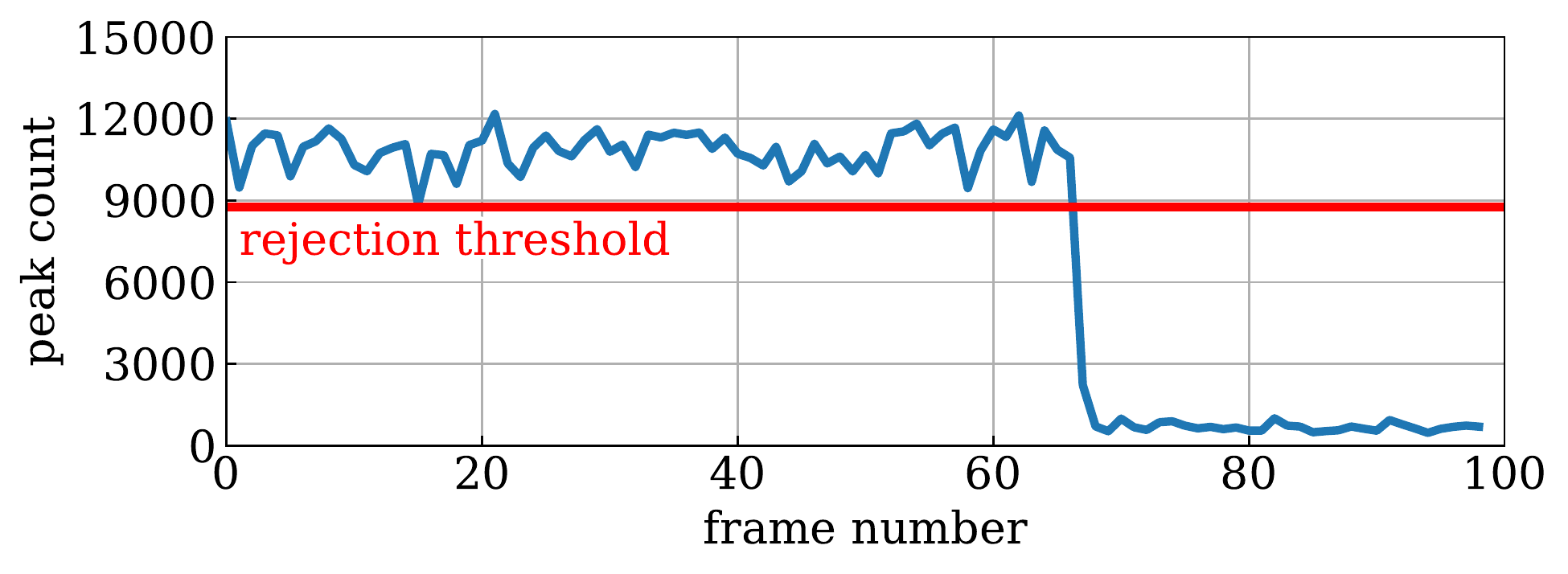}
    \caption{Peak count for all 99 frames of a data cube of HIP~116258. The horizontal red line marks the rejection threshold computed according to Section~\ref{sec:frame_selection}. Around frame 70 the observing conditions suddenly become worse and a clear drop in the peak count can be observed.}
    \label{fig:hip_116258_lucky_imaging}
\end{figure}

As explained in the Introduction, a high Strehl ratio is essential for the kernel phase technique in order for the mathematical framework (i.e. the linearization of the Fourier plane phase, cf. Equation~\ref{eqn:phase}) to be valid. Therefore, we select frames with sufficient Strehl ratio based on their peak pixel count. For each data cube, we first compute the median peak count of the 10\% best frames. Then, we reject all frames with a peak count below 75\% of this value. Using this dynamic threshold is better than simply rejecting a fixed fraction of the frames \citep[e.g.][]{law2006} because it can correctly account for a sudden drop in the Strehl ratio like shown in Figure~\ref{fig:hip_116258_lucky_imaging}. Note that we consider the peak pixel count after performing the PSF reconstruction (cf Section~\ref{sec:reconstruction_saturated_pixels}) here.

\subsection{Kernel phase extraction}
\label{sec:kernel_phase_extraction}

\subsubsection{VLT pupil model}
\label{sec:vlt_pupil_model}

\begin{figure}
    \centering
    \includegraphics[width=\columnwidth]{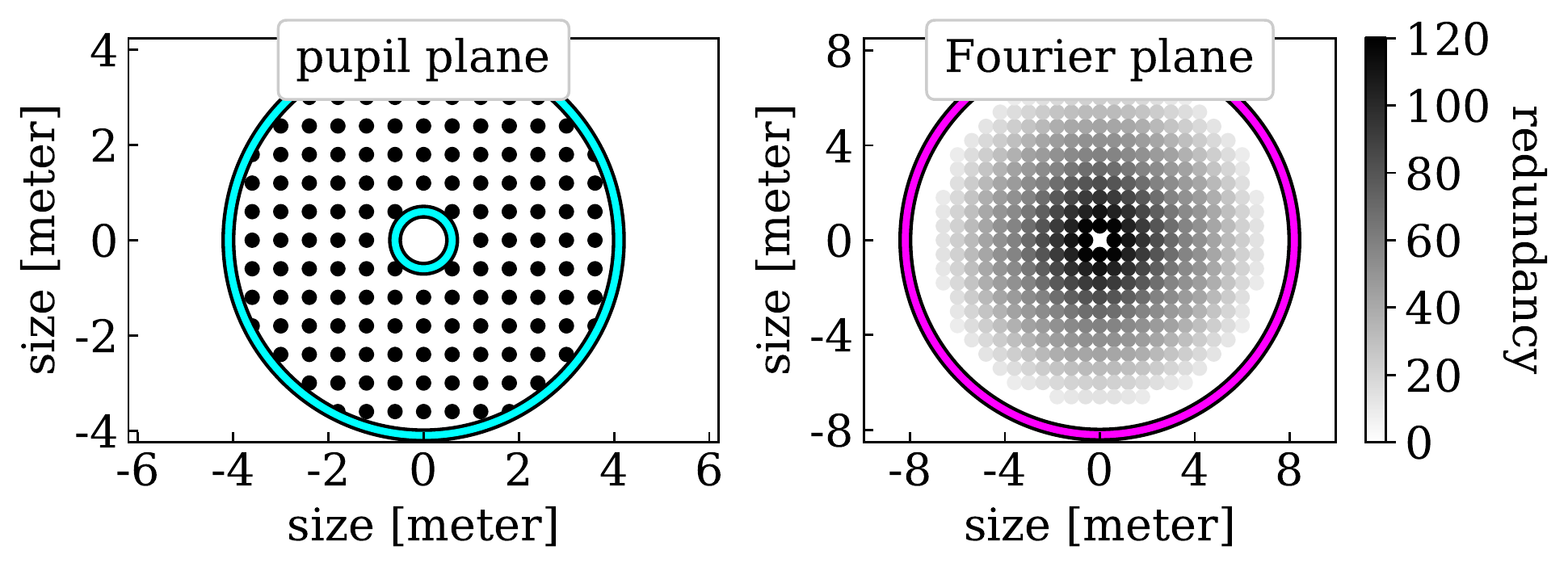}
    \caption{Left panel: our VLT pupil model consisting of 140 \textit{virtual} sub-apertures sampled on a square grid with a pupil plane spacing of $0.6~\text{m}$. The cyan circles show the size of the primary mirror and the central obscuration. Right panel: Fourier plane coverage of the same pupil model. The magenta circle shows the region of support permitted by the pupil geometry in the left panel. Since the Fourier transform is symmetric we only use the phase measured in one half-plane. Note that only these Fourier plane positions within $7.0~\text{m}$ from the origin (i.e. these which do not suffer from low power, cf. Section~\ref{sec:xara}) are shown.}
    \label{fig:vlt_pupil_model}
\end{figure}

In order to extract the kernel phase from VLT/NACO data we first need to construct a model for the VLT pupil (i.e. split the primary mirror into an interferometric array of \textit{virtual} sub-apertures). We sample 140 \textit{virtual} sub-apertures on a square grid with a pupil plane spacing of $0.6~\text{m}$, which is approximately half the Nyquist sampling of $\lambda/\alpha \approx 0.3~\text{m}$, where $\lambda = 3.8~\text{\textmu m}$ is the observing wavelength and $\alpha = 2.610~\text{arcsec}$ is the image size (96 pixels). Our VLT pupil model is shown in the left panel of Figure~\ref{fig:vlt_pupil_model} and based on an $8.2~\text{m}$ primary mirror with a $1.2~\text{m}$ central obscuration. Another advantage of kernel phase over sparse aperture masking is the dense Fourier plane coverage which is shown in the right panel of Figure~\ref{fig:vlt_pupil_model}.

\subsubsection{XARA}
\label{sec:xara}

The extraction of the Fourier plane phase and the computation of the kernel phase relies on a python package called XARA\footnote{\url{https://github.com/fmartinache/xara}} \citep[eXtreme Angular Resolution Astronomy,][]{martinache2010,martinache2013}. XARA has been designed to process data produced by multiple instruments assuming that the images comply to the kernel phase analysis requirements of proper sampling, high-Strehl (boosted by our frame selection procedure described in Section~\ref{sec:frame_selection}), and non-saturation (restored by the procedure described in Section~\ref{sec:reconstruction_saturated_pixels}). The discrete achromatic representation of the VLT aperture (i.e. our pupil model) is used by XARA to compute the phase transfer matrix $\bm{A}$ and the associated left kernel operator $\bm{K}$ via a singular value decomposition of $\bm{A}$.

\begin{figure}
    \centering
    \includegraphics[width=\columnwidth]{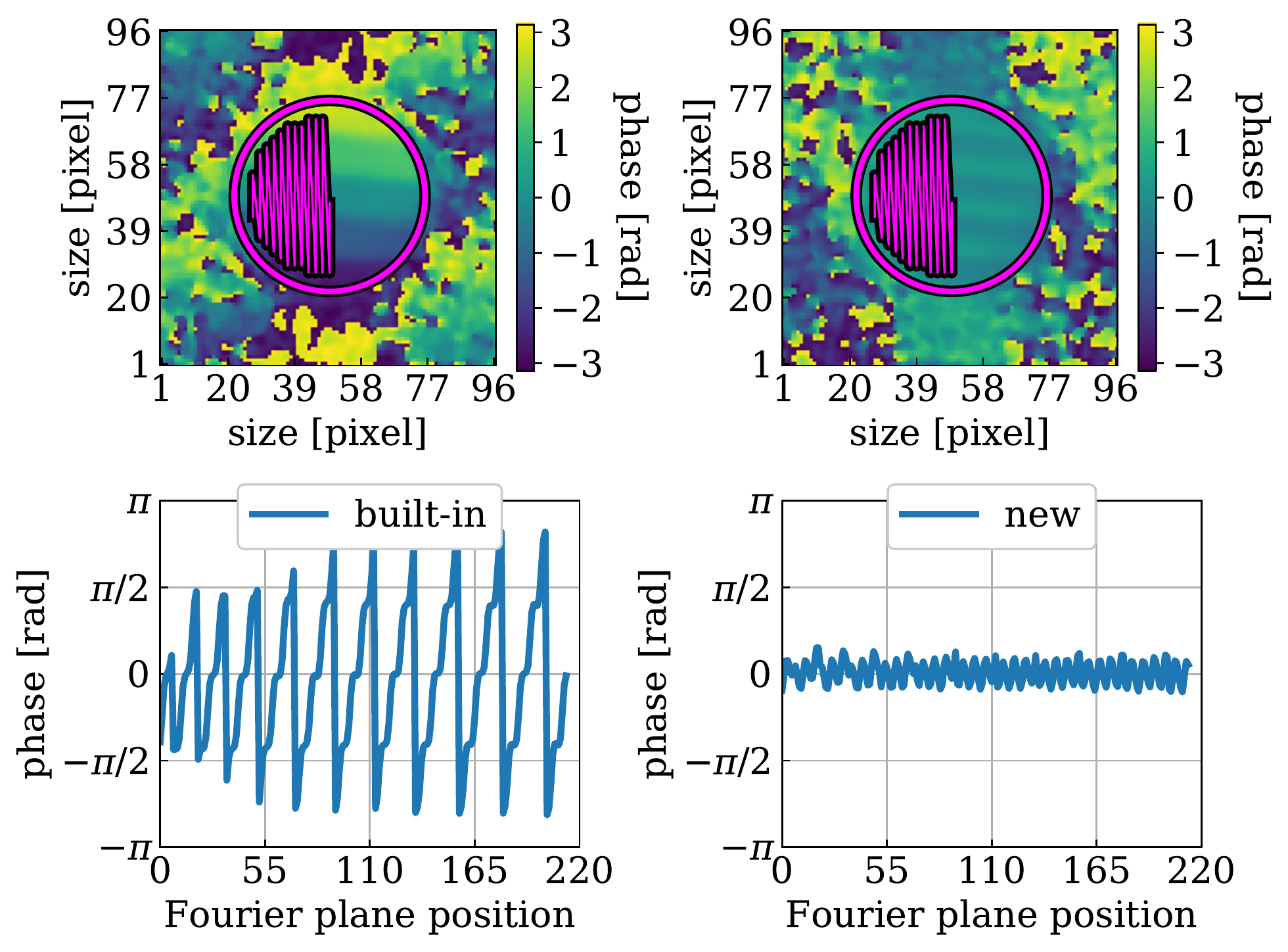}
    \caption{Left panels: Fourier plane phase of the median frame of a data cube of TYC~6849~1795~1 (resolved and bright binary) after imperfect re-centering of the frames (top). The phase is flat in the center, but there is an overall phase ramp from bottom to top caused by the resolved and bright companion. Median Fourier plane phase at the spatial frequencies of our pupil model (bottom). Right panels: same as in the left panels, but after proper re-centering of the frames. The residual Fourier plane phase is of considerably reduced amplitude and can be properly assembled to form meaningful kernel phases. The magenta circles and lines represent the same as in Figure~\ref{fig:hip_47425_fourier_phase}.}
    \label{fig:tyc_6849_1795_1_fourier_phase}
\end{figure}

With the added knowledge of the detector pixel scale and the observing wavelength, the discrete model is scaled so that the Fourier plane phase at the expected $(u,v)$ coordinates can be extracted by a discrete Fourier transform. For the small aberration hypothesis to remain valid, the data must be properly centered prior to the Fourier transform. Failure to do so will leave a residual Fourier plane phase ramp that can wrap and lead to meaningless kernel phases (cf. left panels of Figure~\ref{fig:tyc_6849_1795_1_fourier_phase}). XARA offers several centering algorithms. It is crucial to carefully choose from the available options depending on the requirements coming from the data. For our extensive ground-based data set for example, we find that minimizing directly the Fourier plane phase which is extracted by XARA is most robust and the offered sub-pixel re-centering is very valuable (cf. right panels of Figure~\ref{fig:tyc_6849_1795_1_fourier_phase}) due to an increased level of pupil plane phase noise from the atmosphere and the bright background (if compared to space-borne data).

Moreover, virtual baselines near the outer edge of the Fourier coverage suffer from low power as they are only supported by very few baselines, i.e. have small redundancy. The phase measured for these baselines is systematically noisier and needs to be excluded from the model to prevent the noise to propagate into the estimation of all kernel phases. This can be achieved using the baseline filtering option implemented in XARA. In our case, baselines of length greater than $7.0~\text{m}$ and the corresponding rows of $\bm{A}$ are eliminated prior to the computation of $\bm{K}$. Some of the theoretically available kernel phases are lost but the remaining kernel phases can nevertheless be used just like for the complete model.

Finally, to limit the impact of readout noise in regions of the image where little signal is present, frames are windowed by a super-Gaussian ($g(r) = \exp{-(r/r_0)^4}$) with a radius $r_0 = 25$ pixels, effectively limiting our field of view to $\sim1000~\text{mas}$. Note that Section~\ref{sec:windowing_correction} will further comment on the effect of this window and how it can affect contrast estimates for detections at large separations.

\subsubsection{Kernel phase uncertainties}
\label{sec:kernel_phase_uncertainties}

\begin{figure}
    \centering
    \includegraphics[width=\columnwidth]{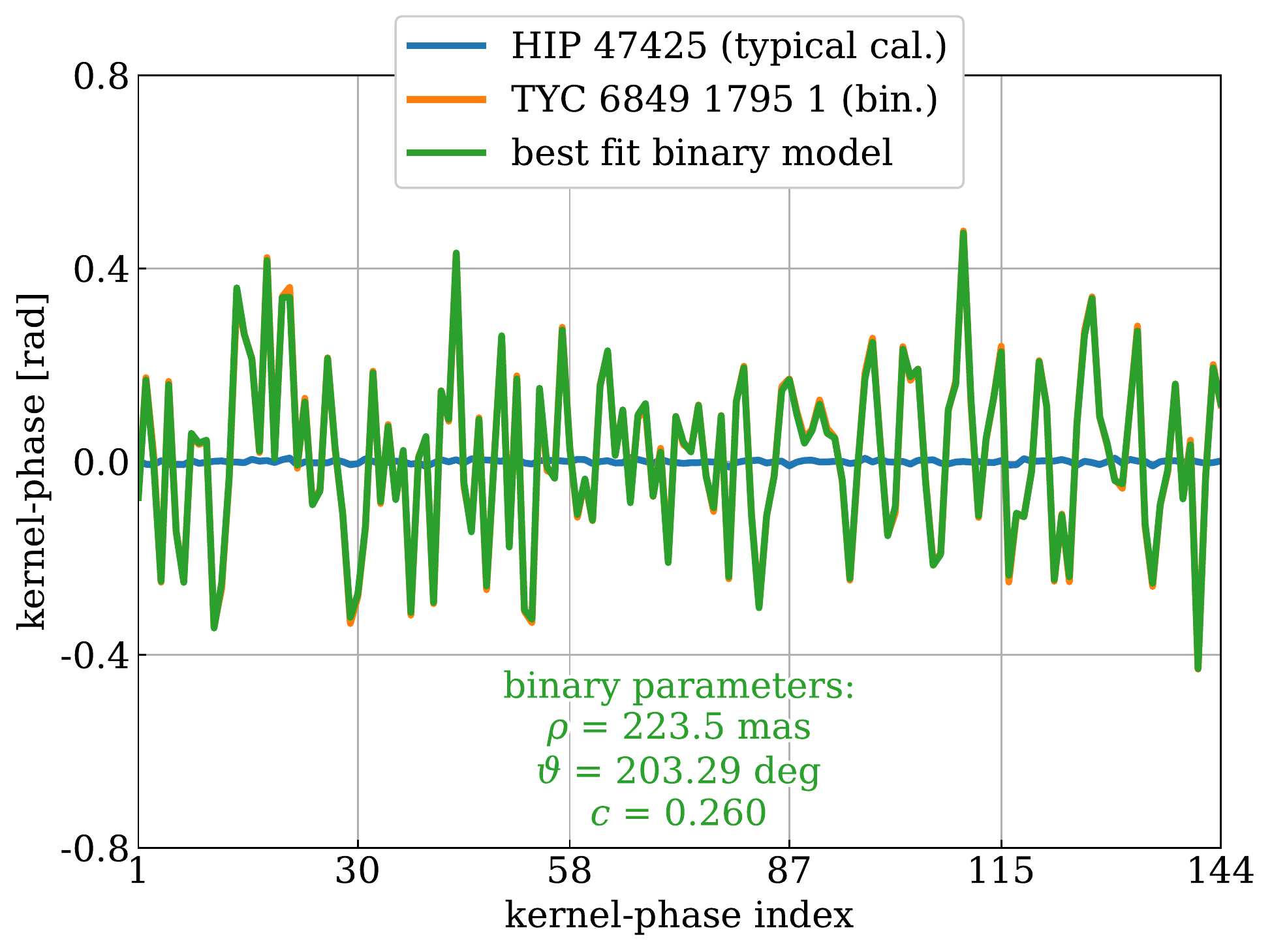}
    \caption{Measured mean kernel phase $\bar{\theta}$ over all data cubes of HIP~47425 (typical calibrator) and TYC~6849~1795~1 (resolved and bright binary) as well as of its best fit binary model $\theta_\text{bin} = \bm{K}\cdot\phi_\text{bin}$ (cf. Section~\ref{sec:binary_model}). Data and model agree very well, so that the green curve overlaps with the orange curve. Note that we normalize each kernel phase by the norm of its corresponding row of $\bm{K}$ and that the raw binary parameters reported here are not corrected for the windowing.}
    \label{fig:hip_47425_kernel_phase_uncertainties}
\end{figure}

For estimating the uncertainties, we compute the kernel phase covariance $\bm{\Sigma}_\theta$ for each frame $\bm{d}$ from its photon count variance $\bm{\Sigma}_d = g\cdot\bm{w}^2\cdot(\bm{d}+\bm{b})$ in units of (photo-electrons)$^2$, where $g$ is the detector gain ($g = 9.8$ for uncorrelated high well depth mode), $\bm{w}$ is the super-Gaussian window, $\bm{d}$ is the cleaned and re-centered frame and $\bm{b}$ is its background (from the simple background subtraction, cf. Section~\ref{sec:dark_flat_background_bad_pixel_correction}). Therefore, we first need to find a linear operator $\bm{B}$ which maps each frame $g\cdot\bm{w}\cdot\bm{d}$ in units of photo-electrons to its kernel phase $\theta$. The linear discrete Fourier transform $\bm{F}$ and the kernel $\bm{K}$ of the pupil model $\bm{R}^{-1}\cdot\bm{A}$ are already linear operators, and the Fourier plane phase $\phi(z)$ (of a complex number $z$) can be approximated as $\mathrm{Im}(z)/|z|$ for small angles. Hence, we compute
\begin{equation}
    \bm{B} = \bm{K}\cdot\frac{\mathrm{Im}(\bm{F})}{\left|\bm{F}\cdot g\cdot\bm{w}\cdot\bm{d}\right|}.
\end{equation}
Note that $\bm{B}\cdot{g}\cdot\bm{w}\cdot\bm{d}$ would be a small-angle approximation for the kernel phase. Then, we obtain an estimate for the kernel phase covariance by propagating the photon count variance according to
\begin{equation}
    \bm{\Sigma}_\theta = \bm{B}\cdot\bm{\Sigma}_d\cdot\bm{B}^T.
\end{equation}

Now, we have a kernel phase $\theta$ and a kernel phase covariance $\bm{\Sigma}_\theta$ for each frame. In order to save computation time for the model fitting (cf. Section~\ref{sec:model_fitting}) we compute a weighted mean $\bar{\theta}$ of the kernel phase for each data cube. Therefore, we first compute the average kernel phase covariance $\bar{\bm{\Sigma}}_\theta$ over all frames $\bm{d}_i$ of a data cube via
\begin{equation}
    \bar{\bm{\Sigma}}_\theta = \left(\sum_i\bm{\Sigma}_{\theta,i}^{-1}\right)^{-1},
\end{equation}
and then the weighted mean $\bar{\theta}$ of the kernel phase (cf. Figure~\ref{fig:hip_47425_kernel_phase_uncertainties}) via
\begin{equation}
    \bar{\theta} = \bar{\bm{\Sigma}}_\theta\cdot\sum_i\bm{\Sigma}_{\theta, i}^{-1}\cdot\theta_i.
\end{equation}
For the rest of this paper, we omit the bar for better readability, i.e.
\begin{align}
    &\bar{\theta} \rightarrow \theta,\\
    &\bar{\bm{\Sigma}}_\theta \rightarrow \bm{\Sigma}_\theta.
\end{align}

Note that this kernel phase covariance model includes the contribution of shot noise only. Any residual calibration errors not taken into account in the following section are therefore expected to increase the reduced $\chi^2$ of any model fitting, potentially to much more than 1.0 in the case of high signal-to-noise data with highly imperfect calibration.

\subsection{Kernel phase calibration}
\label{sec:kernel_phase_calibration}

Under perfect conditions the closure phase of a point-symmetric source, such as an unresolved star, is zero \citep[e.g.][]{monnier2007}. The same holds for the kernel phase, which is a generalization of the closure phase \citep[e.g.][]{ireland2016}. Practically however, one is limited by systematic errors caused by third order phase residuals \citep[e.g.][]{ireland2013} and even point-symmetric sources have non-zero kernel phase.

For this reason, calibration is of fundamental importance when analyzing interferometric measurables (like closure or kernel phase). The systematic errors are expected to be quasi-static \citep[e.g.][]{ireland2013}, i.e. slowly varying with time, so that the kernel phase of a well-known point source measured close in time to that of the science target can serve as a calibrator. The simplest calibration technique would be to subtract the kernel phase of a well-known point source from that of the science target. This technique was for example used successfully in \citet{martinache2010}, but here we want to go beyond this approach.

We use principal component analysis in the framework of a Karhunen-Lo\`eve decomposition \citep{soummer2012, pueyo2016} in order to subtract the statistically most significant phase residuals of the calibrator kernel phase from that of the science target. Note that the following technique is new, but very similar to the POISE observables in \citet{ireland2013}. We start by computing the covariance matrix $\bm{E}_\text{RR}$ of the kernel phase $\theta_{\text{cal},i}$ of all calibrator data cubes $i$ via
\begin{equation}
    \bm{E}_{\text{RR},(i,j)} = \theta_{\text{cal},i}^T\cdot\theta_{\text{cal},j}.
\end{equation}
Then, we compute an eigendecomposition of this matrix in order to obtain its sorted (in descending order) eigenvalues $w_k$ and eigenvectors $v_k$. Finally, we compute the Karhunen-Lo\`eve transform $\bm{Z}$ of shape (number of kernel phases, number of calibrator data cubes) via
\begin{equation}
    \bm{Z}_{(n,k)} = \frac{1}{\sqrt{w_k}}\sum_{p}v_k^p\cdot\theta_{\text{cal},p}^n,
\end{equation}
where $v_k^p$ is the p-th component of the k-th eigenvector of $\bm{E}_\text{RR}$ and $\theta_{\text{cal},p}^n$ is the n-th kernel phase of the p-th calibrator data cube.

From the Karhunen-Lo\`eve transform $\bm{Z}$ we obtain a projection matrix $\bm{P}$ via
\begin{equation}
    \bm{P} = \bm{I}-\bm{Z}'\cdot\bm{Z}'^T,
\end{equation}
where $\bm{I}$ is the identity matrix and $\bm{Z}'$ is obtained from the first $K_\text{klip}$ columns of $\bm{Z}$. $K_\text{klip}$ is an integer representing the order of the correction, i.e. how many eigencomponents of the calibrator kernel phase should be corrected for. The projection matrix $\bm{P}$ is of shape (number of kernel phases, number of kernel phases), but it has $K_\text{klip}$ zero eigenvalues by construction. In order to properly represent the dimensions we compute another eigendecomposition of $\bm{P}$ and obtain a new projection matrix $\bm{P}'$, whose columns are those eigenvectors of $\bm{P}$ which correspond to non-zero eigenvalues. The projection matrix $\bm{P}'$ is of shape (number of ``good'' kernel phases, number of kernel phases), where ``good'' means statistically independent of systematic errors, and can be used to project the measured kernel phase $\theta$ and its covariance $\bm{\Sigma}_\theta$ into a sub-space of dimension (number of ``good'' kernel phases), which is more robust with respect to quasi-static errors, via
\begin{align}
    &\theta' = \bm{P}'\cdot\theta,\\
    &\bm{\Sigma}_\theta' = \bm{P}'\cdot\bm{\Sigma}_\theta\cdot\bm{P}'^T.
\end{align}
For the rest of this paper, we omit the prime for better readability, i.e.
\begin{align}
    &\theta' \rightarrow \theta,\\
    &\bm{\Sigma}_\theta' \rightarrow \bm{\Sigma}_\theta.
\end{align}

\subsection{Model fitting}
\label{sec:model_fitting}

From Equations~\ref{eqn:kernel_phase_begin}--\ref{eqn:kernel_phase_end} it becomes clear that the measured kernel phase $\theta$ directly represents the kernel phase intrinsic to the observed object $\theta_\text{obj}$. Hence, we can infer information about the spatial structure of the observed object by fitting models for $\theta_\text{obj} = \bm{K}\cdot\phi_\text{obj}$ to $\theta$.

\subsubsection{Binary model}
\label{sec:binary_model}

In order to search for companion candidates we use the binary model
\begin{equation}
    r_\text{bin}\cdot{e}^{i\phi_\text{bin}} = 1+c\cdot\exp\left(-2\pi i\cdot\left(\frac{\Delta_\text{RA}\cdot{u}}{\lambda}+\frac{\Delta_\text{DEC}\cdot{v}}{\lambda}\right)\right),
\end{equation}
where $c$ is the contrast ratio between secondary and primary, $u$ and $v$ are the coordinates of the sampled Fourier plane positions (i.e. the spatial frequencies of the pupil model), $\lambda$ is the observing wavelength and
\begin{align}
    &\Delta_\text{RA} = -\rho\cdot\sin(\vartheta-\vartheta_0),\\
    &\Delta_\text{DEC} = \rho\cdot\cos(\vartheta-\vartheta_0),
\end{align}
where $\rho$ is the angular separation between primary and secondary, $\vartheta$ is the position angle of the secondary with respect to the primary and $\vartheta_0$ is the detector position angle during the observation. Figure~\ref{fig:hip_47425_kernel_phase_uncertainties} shows the best fit binary model for the measured kernel phase of TYC~6849~1795~1 (resolved and bright binary).

\subsubsection{Uncertainties from photon noise}
\label{sec:uncertainties_from_photon_noise}

Using the kernel phase covariance $\bm{\Sigma}_\theta$ estimated from photon noise according to Section~\ref{sec:kernel_phase_uncertainties} we compute the best fit contrast $c_\text{fit}$ and its uncertainty $\sigma_{c_\text{fit}}$ for the binary model $\theta_\text{bin} = \bm{K}\cdot\phi_\text{bin}$ on each position of a discrete $500\times500~\text{mas}$ square grid with spacing $13.595~\text{mas}$ (which is half the detector pixel scale of CONICA). In some cases, where we suspect a companion candidate at a larger angular separation, we also extend the grid to $1000\times1000~\text{mas}$.

In the high-contrast regime (where $c \ll 1$), the phase $\phi_\text{bin}$ is approximately proportional to the contrast $c$ of the binary model, so is its kernel phase $\theta_\text{bin}$ (because $\bm{K}$ is a linear operator). Hence, the $\chi^2$ of the binary model $\chi_\text{bin}^2$ can be approximated as
\begin{equation}
    \chi_\text{bin}^2 = (\Theta-c\cdot\Theta_\text{bin,ref})^T\cdot\bm{\Sigma}_\Theta^{-1}\cdot(\Theta-c\cdot\Theta_\text{bin,ref}),
\end{equation}
where $\Theta$ and $\Theta_\text{bin,ref}$ are vertical stacks of the kernel phase $\theta_i$ and the reference binary model $\theta_{\text{bin,ref},i}$ of each data cube~$i$ and $\bm{\Sigma}_\Theta^{-1}$ is a block-diagonal matrix whose diagonal elements are the inverse kernel phase covariances $\bm{\Sigma}_{\theta,i}^{-1}$ of each data cube~$i$, i.e.
\begin{equation}
    \Theta = \begin{pmatrix} \theta_1\\ \theta_2\\ \vdots \end{pmatrix}, \qquad \bm{\Sigma}_\Theta^{-1} = \begin{pmatrix} \bm{\Sigma}_{\theta,1}^{-1} & 0 & \cdots\\ 0 & \bm{\Sigma}_{\theta,2}^{-1} & \cdots\\ \vdots & \vdots & \ddots \end{pmatrix}.
\end{equation}
The reference binary model $\theta_\text{bin,ref}$ is the binary model $\theta_\text{bin}$ evaluated for and normalized by a reference contrast $c_\text{ref} = 0.001$, i.e.
\begin{equation}
    \theta_\text{bin,ref} = \frac{\theta_\text{bin}(c = c_\text{ref})}{c_\text{ref}}.
\end{equation}
Finally, we obtain the log-likelihood $\ln{L}$ for the binary model $\theta_\text{bin}$ as
\begin{equation}
    \ln{L} = -\frac{1}{2}\chi_\text{bin}^2.
\end{equation}

The best fit contrast $c_\text{fit}$ for the binary model $\theta_\text{bin}$ is then obtained by maximizing $\ln{L}$ for each grid position, i.e.
\begin{align}
    \left.\frac{\partial}{\partial{c}}\ln{L}\right|_{c_\text{fit}} &= 0,\\
    \Rightarrow c_\text{fit}                                                  &= \frac{\Theta_\text{bin,ref}^T\cdot\bm{\Sigma}_\Theta^{-1}\cdot\Theta}{\Theta_\text{bin,ref}^T\cdot\bm{\Sigma}_\Theta^{-1}\cdot\Theta_\text{bin,ref}},
\end{align}
and its uncertainty is the square root of its variance, i.e.
\begin{equation}
    \sigma_{c_\text{fit}} = \frac{1}{\sqrt{\Theta_\text{bin,ref}^T\cdot\bm{\Sigma}_\Theta^{-1}\cdot\Theta_\text{bin,ref}}}.
\end{equation}
Finally, the detection significance based on photon noise $\text{SNR}_\text{ph}$ is computed for each grid position as
\begin{equation}
    \text{SNR}_\text{ph} = \frac{c_\text{fit}}{\sigma_\text{ph}} = \frac{c_\text{fit}}{\sigma_{c_\text{fit}}\cdot\sqrt{\chi_\text{r,bin,min}^2}},
\end{equation}
where we scale the uncertainty of the best fit contrast $\sigma_{c_\text{fit}}$ by the square root of the minimal reduced $\chi^2$ of the binary model of the entire grid ($\chi_\text{r,bin,min}^2$). Assuming that kernel phase is proportional to contrast, this is equivalent to scaling the kernel phase covariance $\bm{\Sigma}_\theta$ so that the minimal reduced $\chi^2$ is 1.0. This step is necessary because kernel phase is still affected by third (or higher) order pupil plane phase noise (cf. Equations~\ref{eqn:kernel_phase_begin}--\ref{eqn:kernel_phase_end}), so that the uncertainties from photon noise $\sigma_{c_\text{fit}}$ significantly underestimate the true errors. Note that there can be various sources of higher order phase noise \citep[e.g.][]{ireland2013}, but studying those in detail is beyond the scope of this paper.

The final parameters $p_\text{fit}$ for the best fit binary model are then obtained from a least squares search which maximizes the log-likelihood $\ln{L}$ of the binary model under varying angular separation, position angle and contrast simultaneously. For the least squares search, we use the grid position with the maximal log-likelihood as prior and restrict the search box for the angular separation $\rho$ to $50~\text{mas} \leq \rho \leq 1000~\text{mas}$.

The uncertainties of the best fit parameters $\sigma_{p_\text{fit}}$ follow from the likelihood function $L$ for Gaussian errors (which are applicable to high confidence detections)
\begin{align}
    \ln{L}(p|x) &= -\frac{1}{2}\chi_\text{bin}^2\\
                &= -\frac{1}{2}(\Theta-\Theta_\text{bin}(p))^T\cdot\bm{\Sigma}_\Theta^{-1}\cdot(\Theta-\Theta_\text{bin}(p)),
\end{align}
where $p$ represents the three-dimensional parameter space of angular separation, position angle and contrast. Differentiating twice and neglecting terms containing second order derivatives of a single parameter yields
\begin{align}
    \bm{H}_{(i,j)} &= \frac{\partial^2}{\partial{p}_i\partial{p}_j}\ln{L}(p|x)\\
                   &\approx \frac{\partial\Theta_\text{bin}(p)}{\partial{p}_i}\cdot\bm{\Sigma}_\Theta^{-1}\cdot\frac{\partial\Theta_\text{bin}(p)}{\partial{p}_j}\\
                   &= -(\bm{J}\cdot\bm{\Sigma}_\Theta^{-1}\cdot\bm{J}^T)_{(i,j)},
\end{align}
where $\bm{J}$ and $\bm{H}$ are the Jacobian and the Hessian matrix of the binary model $\Theta_\text{bin}$. Hence, the covariance matrix of the model parameters $\bm{\Sigma}_p$ can be obtained via
\begin{equation}
    \bm{\Sigma}_p = (-\bm{H})^{-1} = (\bm{J}\cdot\bm{\Sigma}_\Theta^{-1}\cdot\bm{J}^T)^{-1},
\end{equation}
and the uncertainties of the model parameters for the best fit binary model $\sigma_{p_\text{fit}}$ are
\begin{equation}
    \sigma_{p_\text{fit}} = \sqrt{\mathrm{diag}(\bm{\Sigma}_{p_\text{fit}})}.
\end{equation}
We also compute the correlation of the best fit model parameters as
\begin{equation}
    \text{corr} = \frac{\bm{\Sigma}_{p_\text{fit}}}{\sigma_{p_\text{fit}}^T\cdot\sigma_{p_\text{fit}}},
\end{equation}
where $\frac{\cdot}{\cdot}$ denotes element-wise division.

\subsubsection{Empirical uncertainties}
\label{sec:empirical_uncertainties}

\begin{figure*}
    \centering
    \includegraphics[width=\textwidth]{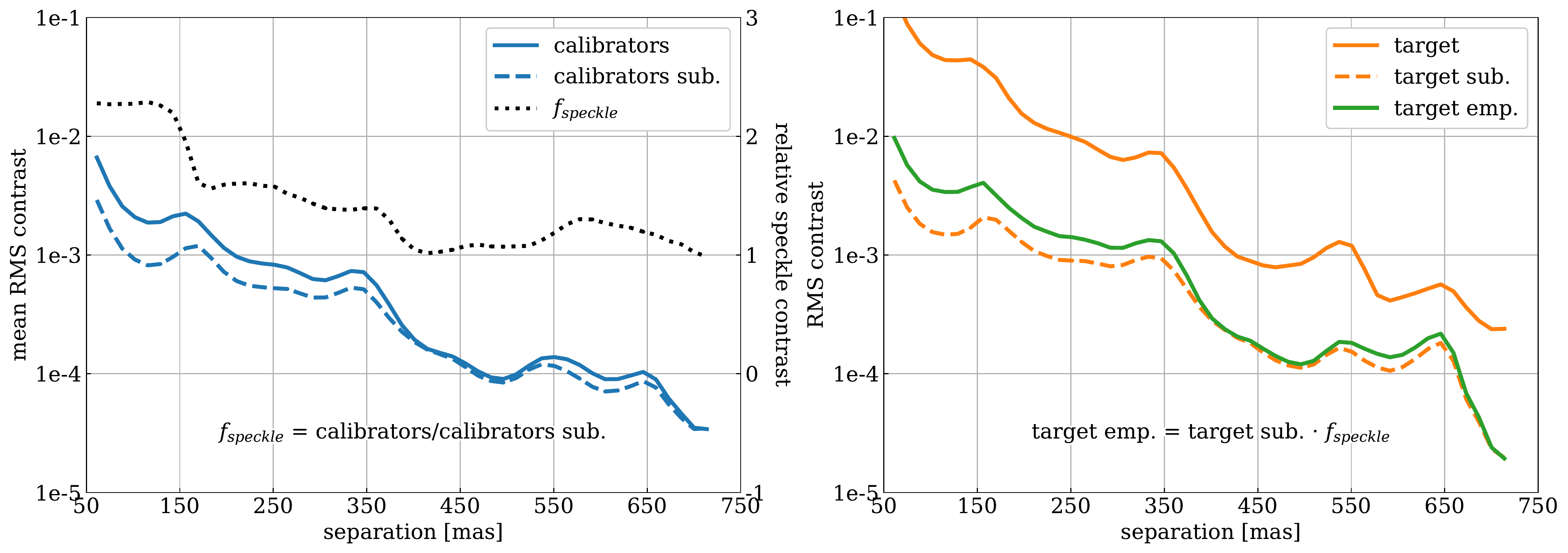}
    \caption{Left panel: mean of the azimuthal average $c_\text{RMS}$ of the RMS best fit contrast $\bm{c}_\text{fit}$ of all non-detections of OB 2 (cf. column ``Det'' of Table~\ref{tab:target_list}) before (solid blue curve) and after (dashed blue curve) subtracting the best fit binary model from the measured kernel phase. The dotted black curve represents the correction factor for the relative contrast of the residual speckle noise $f_\text{speckle}$. Right panel: same as in the left panel, but for HIP~50156 (close binary). The empirical $1\sigma$ detection limit which we use for our analysis (solid green curve) is obtained by multiplying the azimuthal average $c_\text{RMS}^\text{sub}$ of the RMS best fit contrast $\bm{c}_\text{fit}^\text{sub}$ after subtracting the best fit binary model from the measured kernel phase (dashed orange curve) with the correction factor $f_\text{speckle}$.}
    \label{fig:empirical_uncertainties}
\end{figure*}

Using only the uncertainties from photon noise, it is still difficult to distinguish between residual speckle noise (i.e. third order phase noise in the pupil plane) and real detections at small angular separations. This is the case because the data set which we analyze in Section~\ref{sec:results_discussion} is very limited in terms of diversity of calibrator PSFs. For this reason, we use an empirical approach as the primary method to determine whether a detection is real or not.

First, we split our targets into candidate detections and calibrators based on their detection significance from photon noise $\text{SNR}_\text{ph}$ (cf. Section~\ref{sec:close_companion_candidates}). For each of the calibrators, we then compute two contrast curves:
\begin{enumerate}
    \item The azimuthal average $c_\text{RMS}$ of the root mean square (RMS) best fit contrast $\bm{c}_\text{fit}$.
    \item The azimuthal average $c_\text{RMS}^\text{sub}$ of the RMS best fit contrast $\bm{c}_\text{fit}^\text{sub}$ after subtracting the best fit binary model $\theta_\text{bin}$ from the measured kernel phase $\theta$.
\end{enumerate}
Here, the assumption is that the calibrators are single stars, so that the ratio of the two RMS contrast curves computed above, i.e.
\begin{equation}
    f_\text{speckle}(\rho) = \frac{c_\text{RMS}(\rho)}{c_\text{RMS}^\text{sub}(\rho)},
\end{equation}
is a correction factor for the relative contrast of the residual speckle noise. This is illustrated in the left panel of Figure~\ref{fig:empirical_uncertainties}.

For each of the candidate detections, we only compute the azimuthal average $c_\text{RMS}^\text{sub}$ of the RMS best fit contrast $\bm{c}_\text{fit}^\text{sub}$ after subtracting the best fit binary model $\theta_\text{bin}$ (which might or might not be a real detection) from the measured kernel phase $\theta$. Then, we multiply this RMS contrast curve with the mean of the relative speckle contrast $f_\text{speckle}$ of all calibrators, i.e.
\begin{equation}
    \sigma_\text{emp}(\rho) = \bar{f}_\text{speckle}(\rho)\cdot{c}_\text{RMS}^\text{sub}(\rho),
\end{equation}
where the bar denotes the mean, in order to obtain an empirical contrast uncertainty $\sigma_\text{emp}$ as a function of the angular separation $\rho$ for each candidate detection (cf. right panel of Figure~\ref{fig:empirical_uncertainties}). We classify a candidate detection as real if its empirical detection significance $\text{SNR}_\text{emp}$ is above the $5\sigma$ threshold, i.e.
\begin{equation}
    \text{SNR}_\text{emp} = \frac{c_\text{fit}}{\sigma_\text{emp}} > 5.
\end{equation}

Furthermore, we obtain empirically motivated uncertainties on the best fit parameters $p_\text{fit}$ by multiplying the uncertainties from photon noise $\sigma_{p_\text{fit}}$ with the ratio $f_\text{err}$ of the empirical contrast uncertainty $\sigma_\text{emp}$ to the contrast uncertainty from photon noise $\sigma_\text{ph}$ (at the position of the best fit binary model $\theta_\text{bin}$).

The kernel phase analysis tools described in Sections~\ref{sec:kernel_phase_extraction},~\ref{sec:kernel_phase_calibration} and~\ref{sec:model_fitting} are available on GitHub\footnote{\url{https://github.com/kammerje/PyKernel}}. We put a strong focus on applicability to other instruments and an exchangeable kernel phase fits file format.

\section{Results and discussion}
\label{sec:results_discussion}

\subsection{Target list}
\label{sec:target_list}

We test our methods on an archival data set because the kernel phase technique is optimized for detecting companions at much smaller angular separations to their host star than conventional high-contrast imaging techniques (such as ADI and reference star differential imaging, i.e. RDI). Hence, the parameter space that we are looking at is still unexplored. We search the VLT/NACO archive for L' band RDI surveys and decide to analyze program 097.C-0972(A), PI J.~Girard, due to a large number of observed targets and therefore potential calibrators. A target list together with our detections is reported in Table~\ref{tab:target_list}.

\begin{table*}
	\centering
	\caption{Target list grouped by OB for the VLT/NACO program 097.C-0972(A), PI J.~Girard. For each target, we report the spectral type (SpT), the distance ($d$), the apparent K band magnitude (K) and the total integration time after frame selection ($T_\text{int}$). Whether we find any wide (visual) companion candidates, close (kernel phase) candidate detections and real detections is highlighted in columns ``Vis'', ``Can'' and ``Det''. We further report the empirical detection significance for the wide (visual) companion candidates ($\text{SNR}_\text{emp}^\text{vis}$), the detection significance from photon noise for all targets ($\text{SNR}_\text{ph}$) and the empirical detection significance for the close (kernel phase) candidate detections ($\text{SNR}_\text{emp}^\text{can}$).}
	\label{tab:target_list}
	\begin{tabular}{llllllllllll}
		\hline
		OB & Target & SpT & $d$ [pc] & K [mag] & $T_\text{int}$ [s] & Vis & $\text{SNR}_\text{emp}^\text{vis}$ & $\text{SNR}_\text{ph}$ & Can & $\text{SNR}_\text{emp}^\text{can}$ & Det\\
		\hline
		\hline
		\multirow{3}{*}{1} & HIP~68994 & F3/5V & 71.7 & 6.715 & 395.8 & N & -- & 46.6 & Y & 4.7 & N\\
		& HIP~63734 & F7/8V & 54.1 & 6.436 & 389.2 & N & -- & 44.3 & N & -- & N\\
		& HIP~55052 & K7V & 23.7 & 6.808 & 389.2 & N & -- & 45.1 & N & -- & N\\
		\hline
		\multirow{9}{*}{2} & HIP~44722 & K7V & 14.5 & 5.757 & 395.6 & N & -- & 22.1 & N & -- & N\\
		& HD~108767~B & K0V & 26.7 & 6.235 & 310.2 & N & -- & 21.4 & N & -- & N\\
		& HIP~47425 & M3V & 9.6 & 6.056 & 388.8 & N & -- & 32.4 & Y & 1.0 & N\\
		& HIP~50156 & M0.7V & 23.4 & 6.261 & 395.2 & N & -- & 292.7 & Y & 33.2 & Y\\
		& HD~102982 & G3V & 53.2 & 6.605 & 316.6 & N & -- & 23.9 & N & -- & N\\
		& HIP~58029 & G7V & 42.2 & 6.78 & 395.8 & N & -- & 32.9 & Y & 1.4 & N\\
		& HIP~61804 & G3V & 59.2 & 6.869 & 395.8 & N & -- & 27.1 & N & -- & N\\
		& HD~110058 & A0V & 130.0 & 7.583 & 383.4 & N & -- & 30.3 & N & -- & N\\
		& HIP~72053 & G3V & 59.7 & 6.994 & 382.4 & N & -- & 29.4 & N & -- & N\\
		\hline
		\multirow{3}{*}{3} & HIP~58241 & G4V & 35.5 & 6.24 & 256.0 & N & -- & 16.7 & N & -- & N\\
		& TYC~8312~0298~1 & K0II & 804.5 & 6.475 & 162.0 & N & -- & 18.0 & N & -- & N\\
		& HIP~78747 & F5V & 41.1 & 4.859 & 280.8 & N & -- & 22.8 & Y & 2.0 & N\\
		\hline
		\multirow{3}{*}{4} & HIP~37918 & K0IV-V & 34.4 & 6.275 & 389.2 & N & -- & 336.5 & Y & 20.3 & Y\\
		& HIP~36985 & M1.0V & 14.1 & 5.934 & 334.4 & Y & 182.2 & 31.1 & N & -- & N\\
		& TYC~7401~2446~1 & K0V & 42.2 & 6.778 & 117.4 & Y & 195.0 & 14.4 & N & -- & N\\
		\hline
		\multirow{3}{*}{5} & TYC~6849~1795~1 & K5V & 27.6 & 6.911 & 305.4 & Y & 250.1 & 13.3 & N & -- & N\\
		& HIP~92403 & M3.5V & 3.0 & 5.370 & 750.8 & N & -- & 39.1 & Y & 2.5 & N\\
		& HIP~94020~B & K5V & 29.1 & 6.999 & 657.0 & N & -- & 23.9 & N & -- & N\\
		\hline
		\multirow{2}{*}{6} & BDp19~3532 & K0 & 240.2 & 5.842 & 1361.2 & N & -- & -- & N & -- & N\\
		& HIP~108085 & B8IV-V & 64.7 & 3.45 & 401.8 & N & -- & -- & N & -- & N\\
		\hline
		\multirow{2}{*}{7} & HIP~116231 & B9.5III & 53.4 & 4.611 & 285.2 & Y & 4.3 & -- & N & -- & N\\
		& HIP~116258 & K2V & 34.0 & 6.685 & 367.0 & N & -- & -- & N & -- & N\\
		\hline
		8 & HIP~11484 & B9III & 60.4 & 4.392 & 279.6 & N & -- & -- & N & -- & N\\
		\hline
		9 & HIP~3203~B & K5V & 26.5 & 6.834 & 181.6 & N & -- & -- & N & -- & N\\
		\hline
		\multirow{2}{*}{10} & TYC~5835~0469~1 & G8V & 60.9 & 6.997 & 465.0 & Y & 95.8 & -- & N & -- & N\\
		& TYC~9339~2158~1 & K3V & 30.3 & 6.712 & 461.2 & N & -- & -- & N & -- & N\\
		\hline
		\multirow{2}{*}{11} & HIP~7554 & M0V & 22.2 & 6.621 & 637.4 & N & -- & -- & N & -- & N\\
		& HIP~13754 & K2V & 38.6 & 6.883 & 503.8 & N & -- & -- & N & -- & N\\
		\hline
		\multirow{3}{*}{12} & HIP~116384 & K7V & 20.8 & 6.044 & 739.4 & Y & 99.7 & 26.2 & N & -- & N\\
		& HIP~12925 & F8 & 57.1 & 6.52 & 595.4 & N & -- & 24.0 & N & -- & N\\
		& HIP~13008 & F5V & 39.1 & 5.442 & 617.8 & N & -- & 127.4 & Y & N/A & N\\
		\hline
		\multirow{4}{*}{13} & HIP~14555 & M1V & 19.6 & 6.367 & 609.6 & N & -- & 35.1 & N & -- & N\\
		& HIP~20737 & G9.5V & 35.6 & 6.742 & 626.6 & N & -- & 31.1 & N & -- & N\\
		& HIP~22506 & G9V & 50.8 & 6.876 & 620.0 & N & -- & 35.8 & Y & 4.3 & N\\
		& HIP~23362 & B9V & 60.5 & 4.974 & 311.8 & N & -- & 28.7 & N & -- & N\\
		\hline
		\multicolumn{12}{p{0.95\textwidth}}{\textbf{Notes.} OBs 6--11 contain only one or two targets and cannot be analyzed with the kernel phase technique due to a lack of calibrators. Spectral types (SpT), distances ($d$) and apparent K band magnitudes (K) are taken from Simbad \citep{wenger2000}.}\\
	\end{tabular}
\end{table*}

\subsection{Detected companion candidates}
\label{sec:detected_companion_candidates}

\begin{table*}
	\centering
	\caption{Wide companion candidates (CC) detected by visually inspecting the cleaned data (upper section) and close companion candidates detected only by the kernel phase technique (lower section). We estimate the apparent L' band magnitude (L') by adding the contrast ($c$) to the apparent K band magnitude of the host star (K, cf. Table~\ref{tab:target_list}). We report the angular separation ($\rho$) and the position angle ($\vartheta$) of our best fit binary model $\theta_\text{bin}$, the ratio of the empirical errors (which are reported here) to the errors from photon noise ($f_\text{err}$) and the reduced $\chi^2$ of our best fit binary model ($\chi_\text{r,bin}^2$) and the raw kernel phase ($\chi_\text{r,raw}^2$). Whether a detection is new or known is highlighted in column ``New'' and a reference for known detections can be found in column ``Ref''.}
	\label{tab:detected_companion_candidates}
	\begin{tabular}{lllllllllll}
		\hline
		Target & CC & L' [mag] & $c$ [pri/sec] & $\rho$ [mas] & $\vartheta$ [deg] & $f_\text{err}$ & $\chi_\text{r,bin}^2$ & $\chi_\text{r,raw}^2$ & New & Ref\\
		\hline
		\hline
		HIP~36985 & B & $8.553\pm0.005$ & $(8.96\pm0.04)\mathrm{e}{-2}$ & $441.5\pm0.2$ & $133.77\pm0.02$ & 21.66 & 61.6 & 6311.3 & Y & --\\
		TYC~7401~2446~1 & B & $8.096\pm0.005$ & $(2.97\pm0.01)\mathrm{e}{-1}$ & $425.8\pm0.3$ & $89.23\pm0.03$ & 8.10 & 7.1 & 1238.9 & Y & --\\
		TYC~6849~1795~1 & B & $8.363\pm0.004$ & $(2.63\pm0.01)\mathrm{e}{-1}$ & $223.5\pm0.4$ & $203.29\pm0.05$ & 14.29 & 13.0 & 6090.0 & N & G16\\
		HIP~116231 & B & $9.04\pm0.02$ & $(1.69\pm0.03)\mathrm{e}{-2}$ & $874.6\pm0.8$ & $254.70\pm0.05$ & 58.19 & 667.5 & 696.7 & N & S10\\
		TYC~5835~0469~1 & B & $9.396\pm0.003$ & $(1.097\pm0.003)\mathrm{e}{-1}$ & $717.9\pm0.2$ & $37.62\pm0.01$ & 23.56 & 17.1 & 1883.7 & Y & --\\
		HIP~116384 & C & $8.732\pm0.001$ & $(8.412\pm0.008)\mathrm{e}{-2}$ & $842.90\pm0.07$ & $346.614\pm0.004$ & 9.18 & 40.6 & 186.4 & N & M03\\
		\hline
		HIP~50156 & B & $8.17\pm0.03$ & $(1.72\pm0.05)\mathrm{e}{-1}$ & $77.3\pm0.8$ & $338.7\pm0.2$ & 19.75 & 22.1 & 1195.7 & N & B15\\
		HIP~37918 & B & $9.56\pm0.05$ & $(4.9\pm0.2)\mathrm{e}{-2}$ & $122\pm5$ & $9.4\pm0.8$ & 46.55 & 17.3 & 1104.6 & Y & --\\
		\hline
		\multicolumn{11}{p{0.95\textwidth}}{\textbf{Notes.} G16 = \citet{galicher2016}, S10 = \citet{schoeller2010}, M03 = \citet{martin2003}, B15 = \citet{bowler2015}.}\\
	\end{tabular}
\end{table*}

Before we search the targets in Table~\ref{tab:target_list} for close companion candidates, we perform a basic vetting procedure by visually inspecting the cleaned data for wide companion candidates (cf. Section~\ref{sec:wide_companion_candidates}). In the field of view, which is limited to $\sim1~\text{arcsec}$ due to the windowing, we find six wide companion candidates (cf. upper section of Table~\ref{tab:detected_companion_candidates}). Three of them are already known and we classify our detections as confirmed, whereas the other three have not been reported before and therefore are new detections. Note that we correct the contrast of the wide companion candidates for the windowing (cf. Section~\ref{sec:windowing_correction}).

After detecting and subtracting off the signal induced by the wide companion candidates, we use the kernel phase technique in order to search for closer and fainter objects (cf. Section~\ref{sec:close_companion_candidates}). We find two companion candidates with an empirical detection significance above the $5\sigma$ threshold, i.e. $\text{SNR}_\text{emp}^\text{can} > 5$ (cf. lower section of Table~\ref{tab:detected_companion_candidates}). One of them is already known and we classify our detection as confirmed, whereas the other one has not been reported before and therefore is a new detection. For HIP~13008 we note that the empirical detection significance is $9.4\sigma$ when using only HIP~116384 as calibrator, but only $1.9\sigma$ when using HIP~12925 due to high residuals and a very large $f_\text{err}$ correction. Therefore, HIP~12925 seems to be a bad calibrator and we do not report any best fit parameters for HIP~13008 due to a lack of credibility. Follow-up observations are required to confirm the true nature of this object. Also note that OBs 6--11 contain only one or two targets and are not analyzed with the kernel phase technique because the diversity of kernel phase amongst calibrators is essential for our empirical detection method. As there are systematic differences between the individual nights in the measured kernel phase, for this paper we are only analyzing OBs which contain at least two PSF calibrators (observed in the same night). Although this choice was made for simplicity and it might be possible to calibrate targets over longer timescales, this adds significant additional complexity which is beyond the scope of this paper.

From the targets for which we detect neither a wide nor a close companion candidate, we compute a contrast curve (i.e. the detection limit as a function of the angular separation) for the kernel phase technique (cf. Section~\ref{sec:detection_limits}).

\subsubsection{Wide companion candidates}
\label{sec:wide_companion_candidates}

\begin{figure*}
    \centering
    \includegraphics[width=\textwidth]{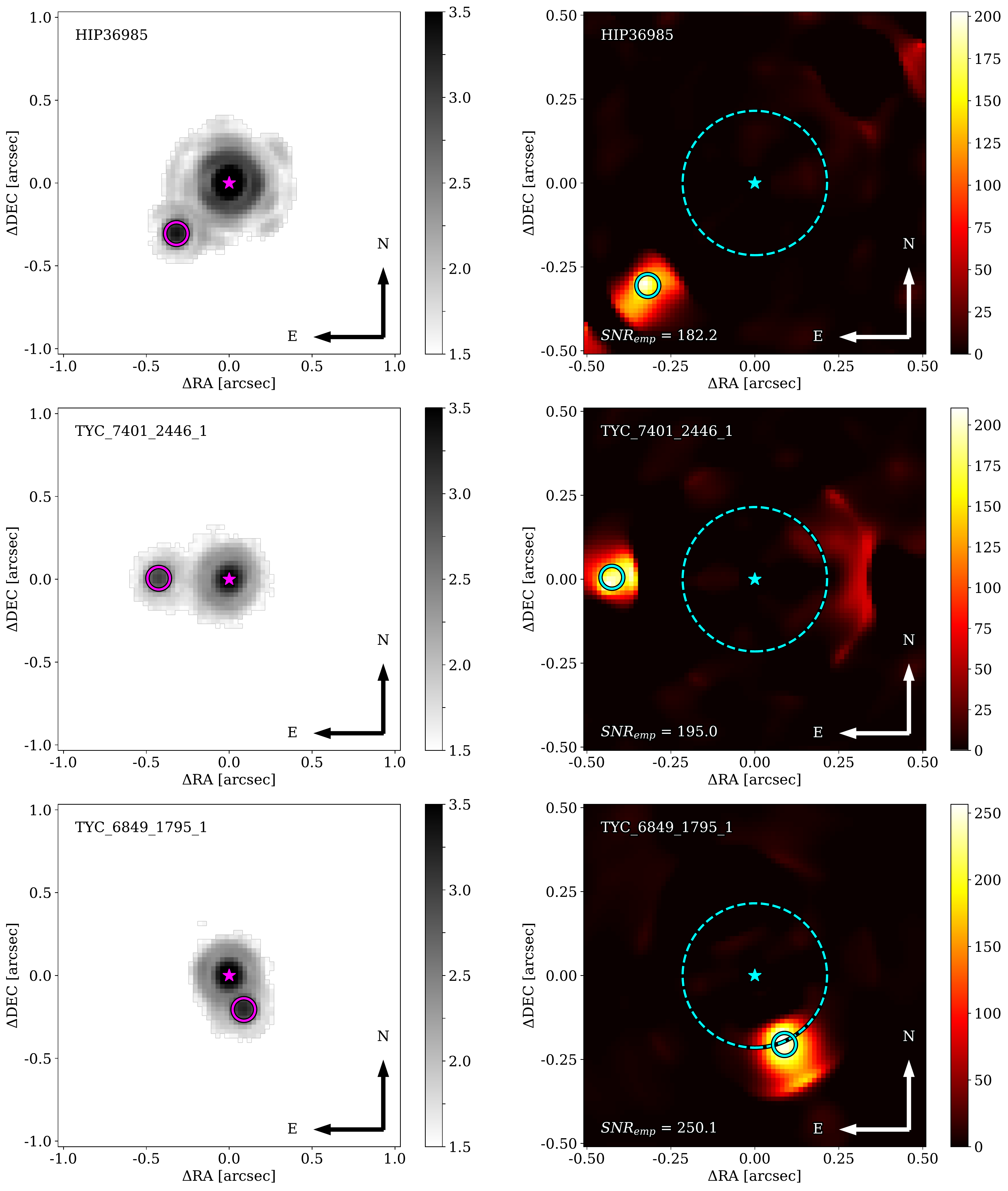}
    \caption{Left panels: median frame of a cleaned data cube of the targets for which we detect a wide companion candidate. The magenta star indicates the position of the host star and the magenta circle indicates the position of the companion candidate, obtained from a least squares fit of the binary model $\theta_\text{bin}$ to the measured kernel phase $\theta$. Note that the color scale is logarithmic, reaching from $1\mathrm{e}{+1.5}$ to $1\mathrm{e}{+3.5}$ pixel counts. Right panels: map of the empirical detection significance $\text{SNR}_\text{emp}^\text{vis}$ (cf. Section~\ref{sec:wide_companion_candidates}) for the same targets as in the left panels. The number in the lower left corner of each panel reports the empirical detection significance at the position of the best fit binary model $\theta_\text{bin}$ (note that this is not necessarily the position with the highest detection significance) and the dashed cyan circle indicates the 99\% threshold of the super-Gaussian window (i.e. the brightness of objects outside this circle is decreased by more than 1\% by the windowing).}
    \label{fig:detections_wide_1}
\end{figure*}

\begin{figure*}
    \centering
    \includegraphics[width=\textwidth]{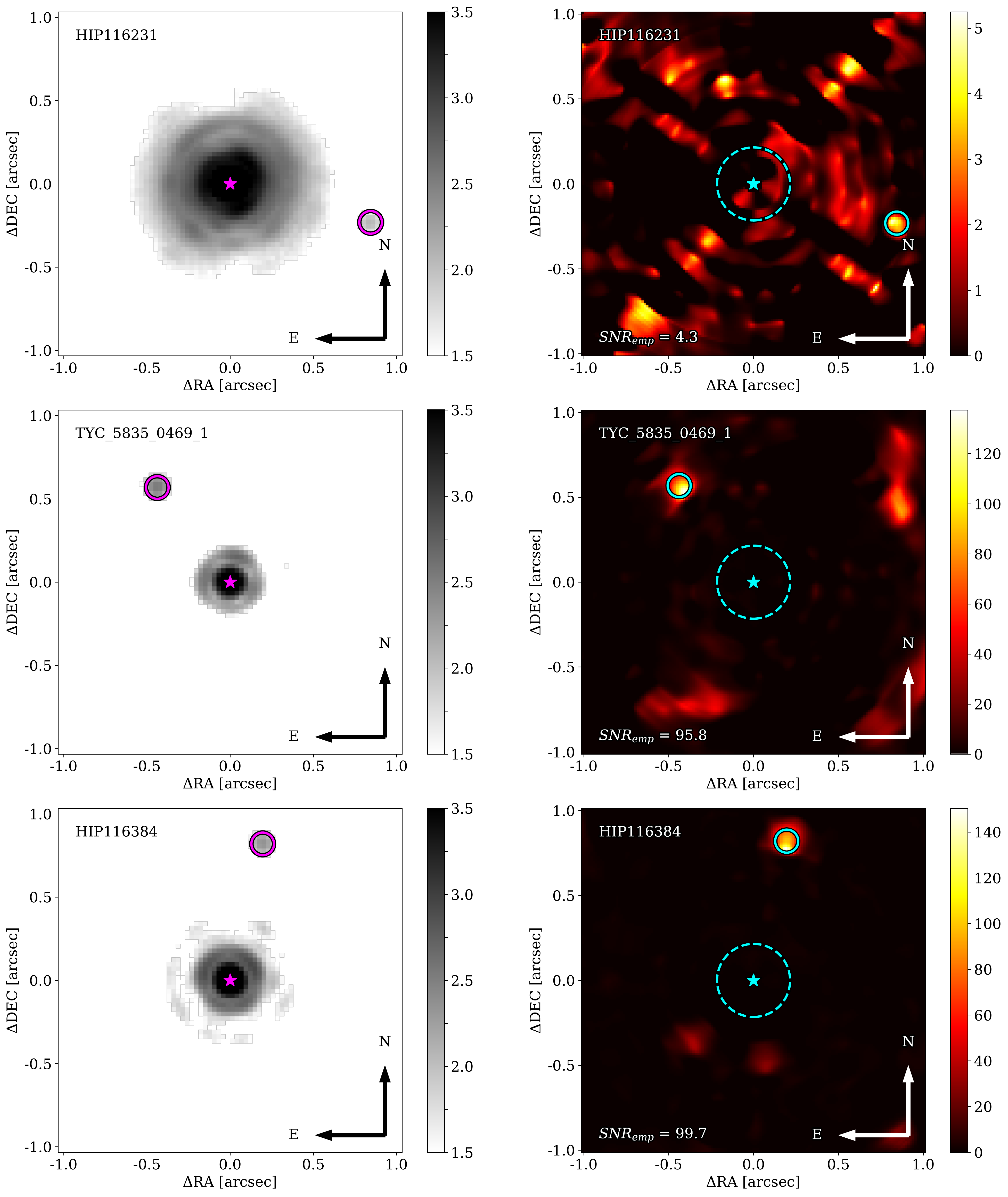}
    \caption{Figure~\ref{fig:detections_wide_1} continued.}
    \label{fig:detections_wide_2}
\end{figure*}

The wide companion candidates reported in the upper section of Table~\ref{tab:detected_companion_candidates} are all detected by visually inspecting the cleaned data. When we find a companion candidate, we use a grid search followed by a least squares search in order to find its best fit binary model $\theta_\text{bin}$ for the measured kernel phase $\theta$. Then, we compute the empirical detection significance $\text{SNR}_\text{emp}^\text{vis}$ for the best fit binary model $\theta_\text{bin}$ (cf. right panels of Figures~\ref{fig:detections_wide_1} and~\ref{fig:detections_wide_2}). This is achieved using a simplification of the empirical detection method (cf. Section~\ref{sec:empirical_uncertainties}). Since the wide companion candidates all have a sufficiently large angular separation (i.e. $\gtrsim200~\text{mas}$) and are sufficiently bright (otherwise we could not detect them by eye), we can skip the use of any calibrators and compute the empirical detection significance $\text{SNR}_\text{emp}^\text{vis}$ as the best fit contrast $\bm{c}_\text{fit}$ divided by the azimuthal average $c_\text{RMS}^\text{sub}$ of the RMS best fit contrast $\bm{c}_\text{fit}^\text{sub}$ after subtracting the best fit binary model $\theta_\text{bin}$ from the measured kernel phase $\theta$. Note that we do not use any Karhunen-Lo\`eve calibration for this step either, i.e. $\theta' = \theta$ (cf. Section~\ref{sec:kernel_phase_calibration}).

Before we search for closer and fainter objects, we subtract the signal induced by the wide companion candidates from the measured kernel phase, i.e.
\begin{equation}
    \theta \rightarrow \theta-\theta_\text{bin},
\end{equation}
so that the measured kernel phase of all targets is free of wide detections. The detected wide companion candidates are shown in the left panels of Figures~\ref{fig:detections_wide_1} and~\ref{fig:detections_wide_2} and are described in more detail in the following paragraphs.

\textbf{HIP~36985~B, TYC~7401~2446~1~B, TYC~5835~0469~1~B.} These objects are new companion candidates which were not reported before. They have L' band contrasts of $2.619\pm0.005^\text{mag}$, $1.318\pm0.004^\text{mag}$ and $2.399\pm0.003^\text{mag}$ respectively, and therefore are candidates for stellar mass companions.
\\
\textbf{TYC~6849~1795~1~B.} This object was already detected in 2005 by \citet{galicher2016} at an angular separation of $\sim220~\text{mas}$, a position angle of $\sim201~\text{deg}$ and a H band contrast of $\sim1.6^\text{mag}$. We find a L' band contrast of $1.450\pm0.004^\text{mag}$ and an angular separation ($223.5\pm0.4~\text{mas}$) and a position angle ($203.29\pm0.05~\text{deg}$) which are in agreement with \citet{galicher2016}, i.e. we can confirm the bound nature of the object.
\\
\textbf{HIP~116231~B.} This object was already detected in 2004 by \citet{schoeller2010} at an angular separation of $641\pm4~\text{mas}$, a position angle of $240.2\pm0.6~\text{deg}$ and a K band contrast of $2.75\pm0.01^\text{mag}$. We find a L' band contrast of $4.43\pm0.02^\text{mag}$, a slightly larger angular separation of $874.6\pm0.8~\text{mas}$ and a slightly different position angle of $254.70\pm0.05~\text{deg}$, but (allowing for orbital motion) we can confirm the bound nature of the object. Note that there is a huge disagreement in the contrast, but a brief look at the raw data from \citet{schoeller2010} shows a significant PSF halo and confirms our result of $\sim4^\text{mag}$.
\\
\textbf{HIP~116384~C.} This object was first detected in 2002 by \citet{martin2003} who found HIP~116384 (GJ~900) to be a triple system with a $510\pm10~\text{mas}$ (HIP~116384~B, $\Delta\text{K} = 1.61\pm0.03^\text{mag}$) and a $760\pm10~\text{mas}$ (HIP~116384~C, $\Delta\text{K} = 2.38\pm0.04^\text{mag}$) component. \citet{lafreniere2007b} resolved the system again in 2004 and 2005, finding HIP~116384~B at an angular separation of $611\pm2~\text{mas}$ and $673\pm2~\text{mas}$ respectively, and HIP~116384~C at an angular separation of $733\pm2~\text{mas}$ and $722\pm2~\text{mas}$ respectively. In the cleaned data, we only find HIP~116384~C at a slightly larger angular separation of $842.90\pm0.07~\text{mas}$, but a position angle ($346.614\pm0.004~\text{deg}$) and a L' band contrast ($2.688\pm0.001^\text{mag}$) which are in agreement with \citet{martin2003} and \citet{lafreniere2007b}, so that we can confirm the bound nature of the object. Looking at the raw data, we also find HIP~116384~B (which is the brighter of the two companions), noticing that it has moved to an angular separation of $\sim1200~\text{mas}$ being too far away in order to be visible in our cleaned data (due to the windowing).

\subsubsection{Close companion candidates}
\label{sec:close_companion_candidates}

\begin{figure}
    \centering
    \includegraphics[width=\columnwidth]{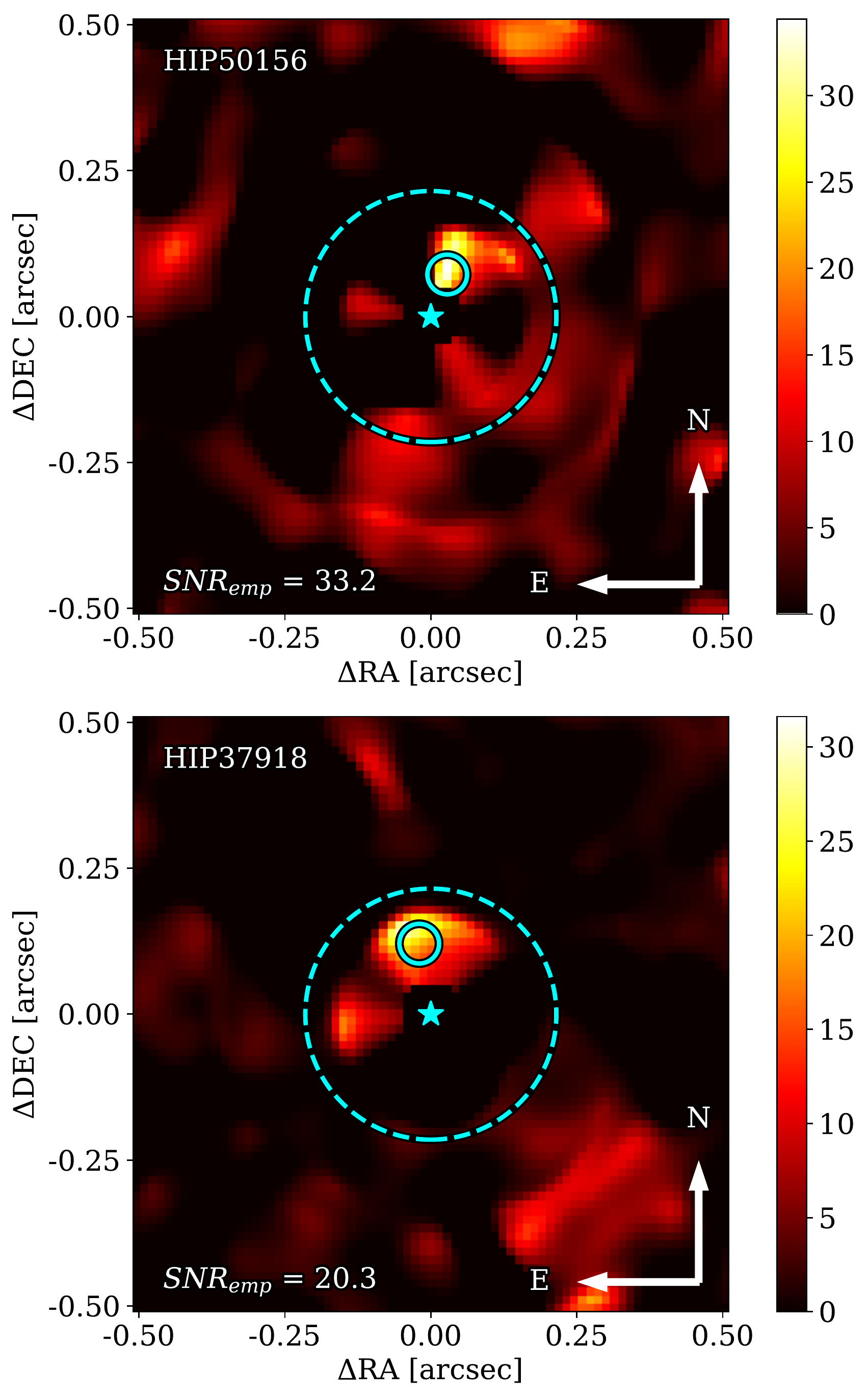}
    \caption{Map of the empirical detection significance $\text{SNR}_\text{emp}^\text{can}$ for the targets for which we detect a close companion candidate. The cyan star indicates the position of the host star and the solid cyan circle indicates the position of the companion candidate, obtained from a least squares fit of the binary model $\theta_\text{bin}$ to the measured kernel phase $\theta$. The number in the lower left corner of each panel reports the empirical detection significance at the position of the best fit binary model $\theta_\text{bin}$ and the dashed cyan circle indicates the 99\% threshold of the super-Gaussian window (like in Figure~\ref{fig:detections_wide_1}).}
    \label{fig:detections_close}
\end{figure}

The close companion candidates reported in the lower section of Table~\ref{tab:detected_companion_candidates} are all detected only by the kernel phase technique. For each target in Table~\ref{tab:target_list}, we use a grid search followed by a least squares search in order to find the best fit binary model $\theta_\text{bin}$ for the measured kernel phase $\theta$. Then, we compute the detection significance from photon noise $\text{SNR}_\text{ph}$ (cf. Section~\ref{sec:uncertainties_from_photon_noise}) at the position of the best fit binary model $\theta_\text{bin}$ from the least squares search. For this step, we always use all other targets which were observed in the same OB as calibrators for the Karhunen-Lo\`eve calibration, fixing $K_\text{klip} = 4$\footnote{For simplicity, we fix $K_\text{klip} = 4$ for all targets and regardless of the number of calibrators. Various testing has shown that subtracting off the four statistically most significant eigencomponents of the kernel phase of the calibrators mostly yields the smallest amount of significant detections, i.e. calibrates the data best. A more rigorous investigation of this relationship is foreseen for a future publication.}. Knowing that the majority of VLT/NACO targets do not have any close companions, we then classify the $\sim1/3$ of the targets with the highest $\text{SNR}_\text{ph}$ in each OB as candidate detections (cf. column ``Can'' of Table~\ref{tab:target_list}) for the next step and the remaining targets as calibrators.

For the next step, we compute the empirical detection significance $\text{SNR}_\text{emp}^\text{can}$ (cf. Section~\ref{sec:empirical_uncertainties}) for each of the candidate detections from the previous step. For this step, we always use all remaining targets which were classified as calibrators in the previous step for the Karhunen-Lo\`eve calibration, again fixing $K_\text{klip} = 4$. If the empirical detection significance is above the $5\sigma$ threshold, i.e. $\text{SNR}_\text{emp}^\text{can} > 5$, we classify the candidate detection as real. If not, we add the candidate detection to the list of calibrators and redo the computation of the empirical detection significance (this time with one calibrator more than before). We repeat this process until all candidate detections are real. The detected close companion candidates are shown in Figure~\ref{fig:detections_close} and are described in more detail in the following paragraphs. Please note that we report the correlation of the best fit parameters in Appendix~\ref{sec:parameter_correlation} and present model-data correlation plots in Appendix~\ref{sec:correlation_plots}.

\textbf{HIP~50156~B}. This object was already detected in 2011 by \citet{bowler2015} at an angular separation of $\sim90~\text{mas}$ and a K band contrast of $\sim1.8^\text{mag}$. Just nine month later, \citet{brandt2014} cannot resolve this companion and report an upper limit of $\sim20~\text{mas}$ for its angular separation. We find HIP~50156~B at an angular separation of $77.3\pm0.8~\text{mas}$ and an L' band contrast of $\sim1.91\pm0.03^\text{mag}$, confirming the detection and notable orbital motion.
\\
\textbf{HIP~37918~B.} This object is a new companion candidate which was not reported before. It has a L' band contrast of $\sim3.29\pm0.05^\text{mag}$, and therefore is a candidate for a stellar mass companion. Furthermore, HIP~37918 ($M \approx 0.98~M_\odot$) is known to have a $\sim23.1~\text{arcsec}$ companion of almost equal mass (HIP~37923, $M \approx 0.95~M_\odot$, \citealt{desidera2006}). Together with our companion candidate, this would make the system triple.

\subsection{Detection limits}
\label{sec:detection_limits}

\begin{figure*}
    \centering
    \includegraphics[width=\textwidth]{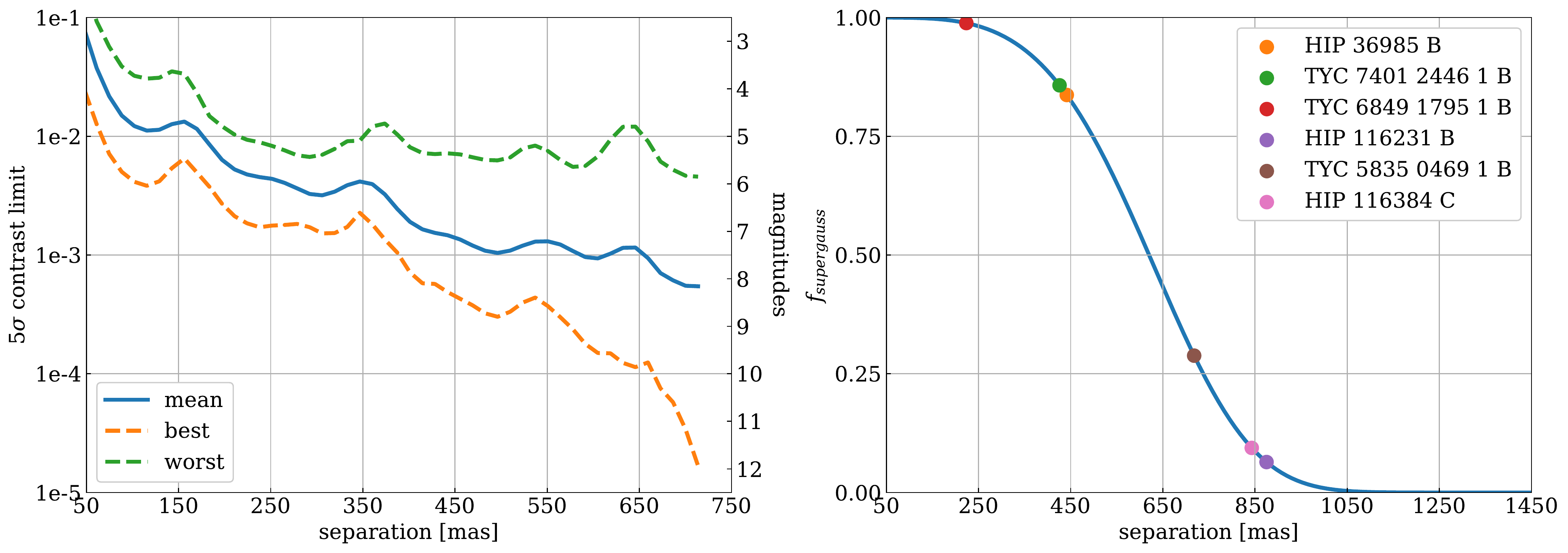}
    \caption{Left panel: $5\sigma$ empirical contrast limit, i.e. RMS contrast curve $c_\text{RMS}(\rho)$ multiplied by 5, for all non-detections (cf. column ``Det'' of Table~\ref{tab:target_list}). Shown are the mean, the best and the worst contrast limit. Right panel: value of the super-Gaussian windowing function depending on the angular separation. The brightness of companions outside of $\sim200~\text{mas}$ is decreased significantly. We use this curve to recover the true contrast of the detected wide (visual) companions (cf. upper section of Table~\ref{tab:detected_companion_candidates}). For reference, their position on this curve is indicated by the circles.}
    \label{fig:contrast_limit_windowing_correction}
\end{figure*}

In Section~\ref{sec:empirical_uncertainties}, we present our empirical approach to find meaningful detection limits for the data analyzed in this paper. Based on this approach, we compute the contrast limit of the kernel phase technique as a function of the angular separation as the azimuthal mean of the RMS best fit contrast $c_\text{RMS}$ of all targets for which we do not detect any companions with the kernel phase technique (i.e. all non-detections, cf. column ``Det'' of Table~\ref{tab:target_list}). Note that we already subtracted off the signal induced by the wide companion candidates. The mean, the best and the worst contrast limit are shown in the left panel of Figure~\ref{fig:contrast_limit_windowing_correction}.

At the small angular separations which are inaccessible by classical high-contrast imaging techniques (i.e. within $\sim200~\text{mas}$ in the L' band), the kernel phase technique achieves contrast limits of $\sim1\mathrm{e}{-2}$. This is not yet deep enough to detect companions in the planetary-mass regime, which would start between $1\mathrm{e}{-3}$ and $1\mathrm{e}{-4}$ for young ($\sim10~\text{Myr}$) gas giants \citep[e.g.][]{bowler2016}. However, our closest detections prove that the resolution which is required to resolve solar-system scales in the nearest star forming regions can be achieved with the kernel phase technique. At larger angular separations, our best contrast limit is comparable with the limits achieved by RDI \citep[e.g.][]{cantalloube2015}. The large spread in the contrast limit comes from the fact that the amplitude of the background noise is nearly the same for all data cubes, whereas the peak value of the PSF varies heavily due to the PSF reconstruction (cf. Section~\ref{sec:reconstruction_saturated_pixels}).

\subsection{Windowing correction}
\label{sec:windowing_correction}

As mentioned in Section~\ref{sec:xara}, we window all frames by a super-Gaussian (with a FWHM of $1240~\text{mas}$) in order to minimize edge effects when computing their Fourier transform. Due to this windowing, the brightness of companions at angular separations $\gtrsim215~\text{mas}$ deviates by more than 1\% from the true value. In order to correct for this effect, we again assume that kernel phase is proportional to contrast in the high-contrast regime, so that we can obtain the true contrast of a companion by dividing its measured contrast (i.e. the best fit contrast from the binary model) by the value of the super-Gaussian windowing function. We are aware that this method has its limits, as each PSF has a spatial extent on the detector and assuming that the entire PSF is multiplied by the same value is an over-simplification of the problem. Nevertheless, this method agrees fairly well with the contrasts which we measure in the cleaned fits files and we use it to correct the contrast of all wide companion candidates (cf. right panel of Figure~\ref{fig:contrast_limit_windowing_correction}). We add an additional contrast correction error in quadrature based on injection-recovery tests to companions wider than $500~\text{mas}$ to account for limitations in this technique.

\section{Conclusions}
\label{sec:conclusions}

We use the kernel phase technique in order to search for close companions at the diffraction limit in an archival VLT/NACO RDI L' band data set. Therefore, we develop our own data reduction pipeline for VLT/NACO data, which performs a basic dark, flat, bad pixel and background (i.e. dither) subtraction, but also reconstructs saturated PSFs in order to reduce their Fourier plane noise. Furthermore, we select frames with sufficiently high Strehl ratio, which is essential for the kernel phase technique as it relies on a linearization of the Fourier plane phase. Then, we use XARA for extracting the kernel phase and improve its re-centering algorithm in the case of resolved and bright companions. Furthermore, we apply a principal component analysis based calibration to the data \citep[i.e. Karhunen-Lo\`eve decomposition,][]{soummer2012} and develop a suite of analytic model fitting algorithms in order to search for point source companions with the kernel phase technique\footnote{\url{https://github.com/kammerje/PyKernel}}.

For the archival data set which we analyze in Section~\ref{sec:results_discussion}, we find that our kernel phase covariance model (which only takes into account shot noise) is not sufficient and significantly underestimates the true errors. This is still the case after calibrating the data, because the diversity of calibrator PSFs is not sufficient. Hence, we develop an empirical method for estimating the relative contrast of the residual speckle noise and finding meaningful detection limits for the data. With this empirical approach, we detect six wide companion candidates by visually inspecting the cleaned data and two close ($\sim80$--$110~\text{mas}$) companion candidates which are detected only by the kernel phase technique. All eight companion candidates lie in the stellar-mass regime and five of them were previously unknown.

In order to reach the planetary-mass regime, a better library of calibrator PSFs is required. Therefore, it is extremely important that the targets and their calibrators are observed as close in time as possible. This becomes very clear from the archival data set which we analyze, where there are in fact multiple calibrators observed in one night, but not close enough in time, so that the kernel phase calibration does not reduce the quasi-static errors satisfyingly. In order to make better use of our prinipal component analysis based calibration, we propose star-hopping sequences of $\sim6$ targets, and to revisit each target at least twice. Star-hopping is an observing strategy for which the instrument (and in particular the AO system) acquisition is only performed once at the beginning of each sequence. Then, one slews (``hops'') from target to target without interrupting the AO system. Furthermore, we aim to examine more extensive Keck data sets where we are hopeful that the significant investment of telescope resources gives adequate calibrator diversity to characterize the systematic errors and possibly use Bayesian Monte-Carlo techniques.

In this paper, we have shown that kernel phase is able to achieve a resolution below the classical diffraction limit of a telescope under good observing conditions (i.e. sufficiently high Strehl ratio). This is of particular interest for future space-based observatories, such as the JWST, as it gives access to an exciting parameter space which could otherwise not be explored due to the limited mirror size (and therefore resolution). Space-based telescopes do not suffer from atmospheric turbulence, what makes the calibration much less challenging than for the ground-based VLT/NACO data \citep[e.g.][]{martinache2010}. Nevertheless, with an optimized observing strategy, kernel phase is also a competitive high-contrast imaging technique from the ground.

The application of kernel phase is of course not limited to imaging telescopes. One concept which aims to push the kernel phase technique towards higher contrasts is the VIKiNG instrument \citep{martinache2018}, which proposes kernel phase nulling interferometry with the VLTI. By combining kernel phase with a high-contrast booster (i.e. a nulling interferometer) it would allow for self-calibrating the observables and achieving a better robustness with respect to residual wavefront errors. This would in turn also be an option to reduce the demanding stability requirements on space-based nulling interferometers, such as the LIFE concept \citep{kammerer2018,quanz2018}, which aims to detect dozens of Earth-like exoplanets in the solar neighborhood.

\section*{Acknowledgements}

MJI was supported by the Australian Research Council Future Fellowship (FT130100235). This project has received funding from the European Research Council (ERC) under the European Union's Horizon 2020 research and innovation program (grant agreement CoG \#683029). JHG gratefully acknowledges support from the Director's Research Funds at the Space Telescope Science Institute. The manuscript was
also substantially improved following helpful comments from an
anonymous referee.

%%%%%%%%%%%%%%%%%%%%%%%%%%%%%%%%%%%%%%%%%%%%%%%%%%



\bibliographystyle{mnras}

%\begin{thebibliography}{99}
%\end{thebibliography}

%%%%%%%%%%%%%%%%%%%%%%%%%%%%%%%%%%%%%%%%%%%%%%%%%%

%%%%%%%%%%%%%%%%% APPENDICES %%%%%%%%%%%%%%%%%%%%%

\appendix

\section{Parameter correlation}
\label{sec:parameter_correlation}

\begin{figure*}
    \centering
    \includegraphics[width=\columnwidth]{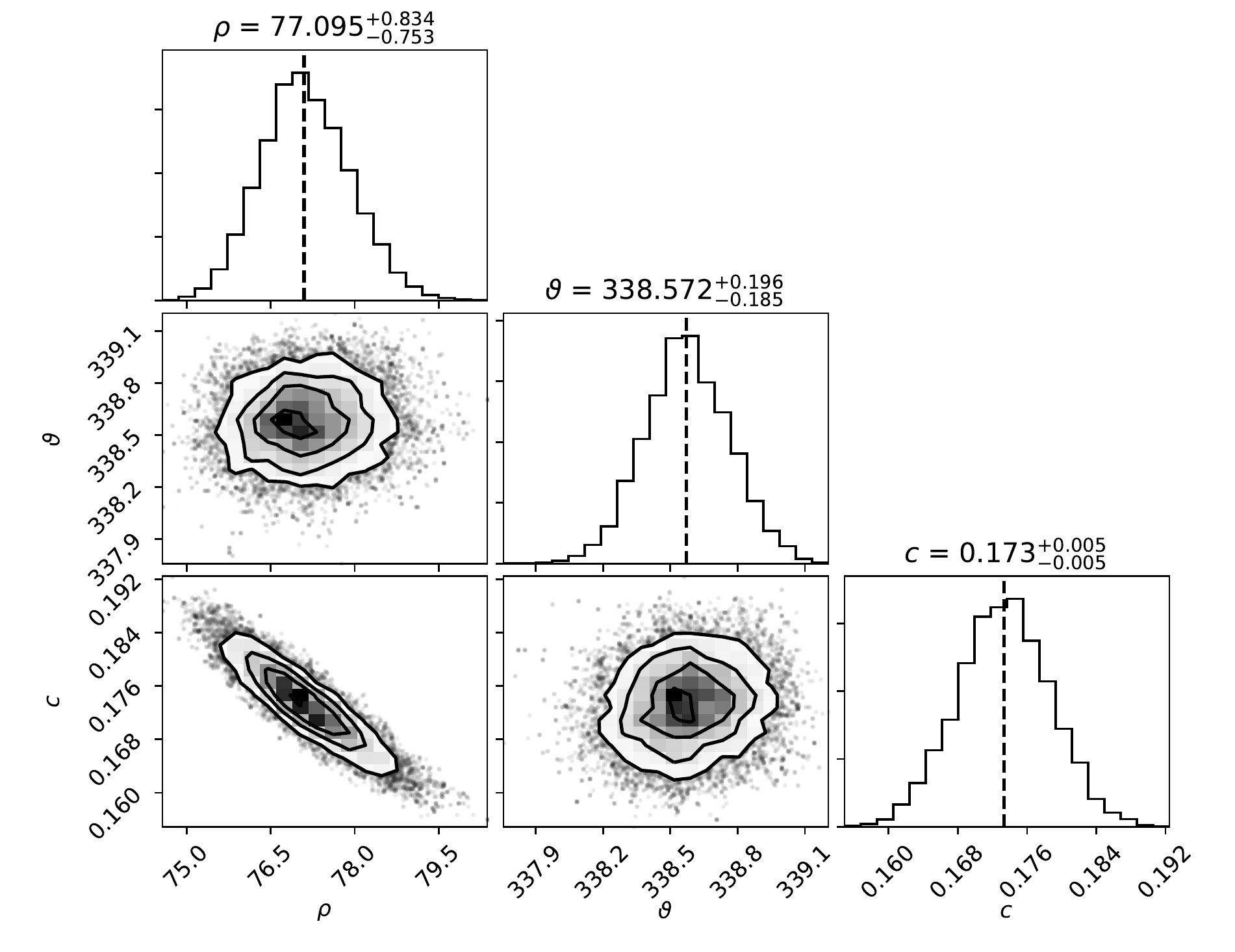}
    \\
    \includegraphics[width=\columnwidth]{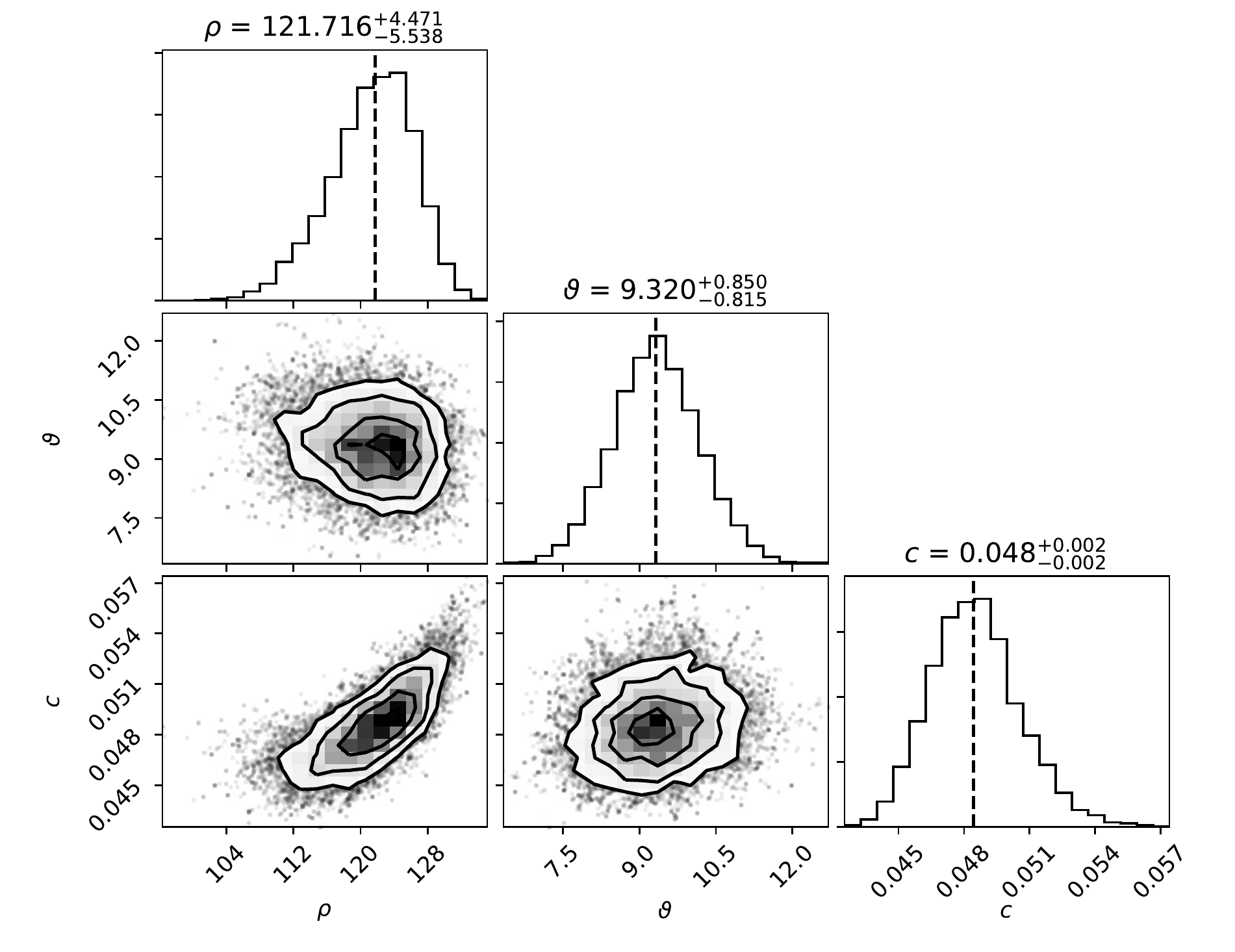}
    \caption{For the targets for which we detect a close companion candidate (i.e. HIP~50156, top and HIP~37918, bottom) we report the correlation of the best fit parameters using a corner plot from \citet{corner}. Here, we use an MCMC technique \citep[emcee,][]{foreman-mackey2013} with six random walkers initialized at the best fit position and a temperature of $f_\text{err}^2$ in order to find the best fit parameters including their correlated uncertainties by maximizing the log-likelihood $\ln{L}$ of the binary model.}
    \label{fig:correlation_plots}
\end{figure*}

\section{Correlation plots}
\label{sec:correlation_plots}

\begin{figure*}
    \centering
    \includegraphics[width=\columnwidth]{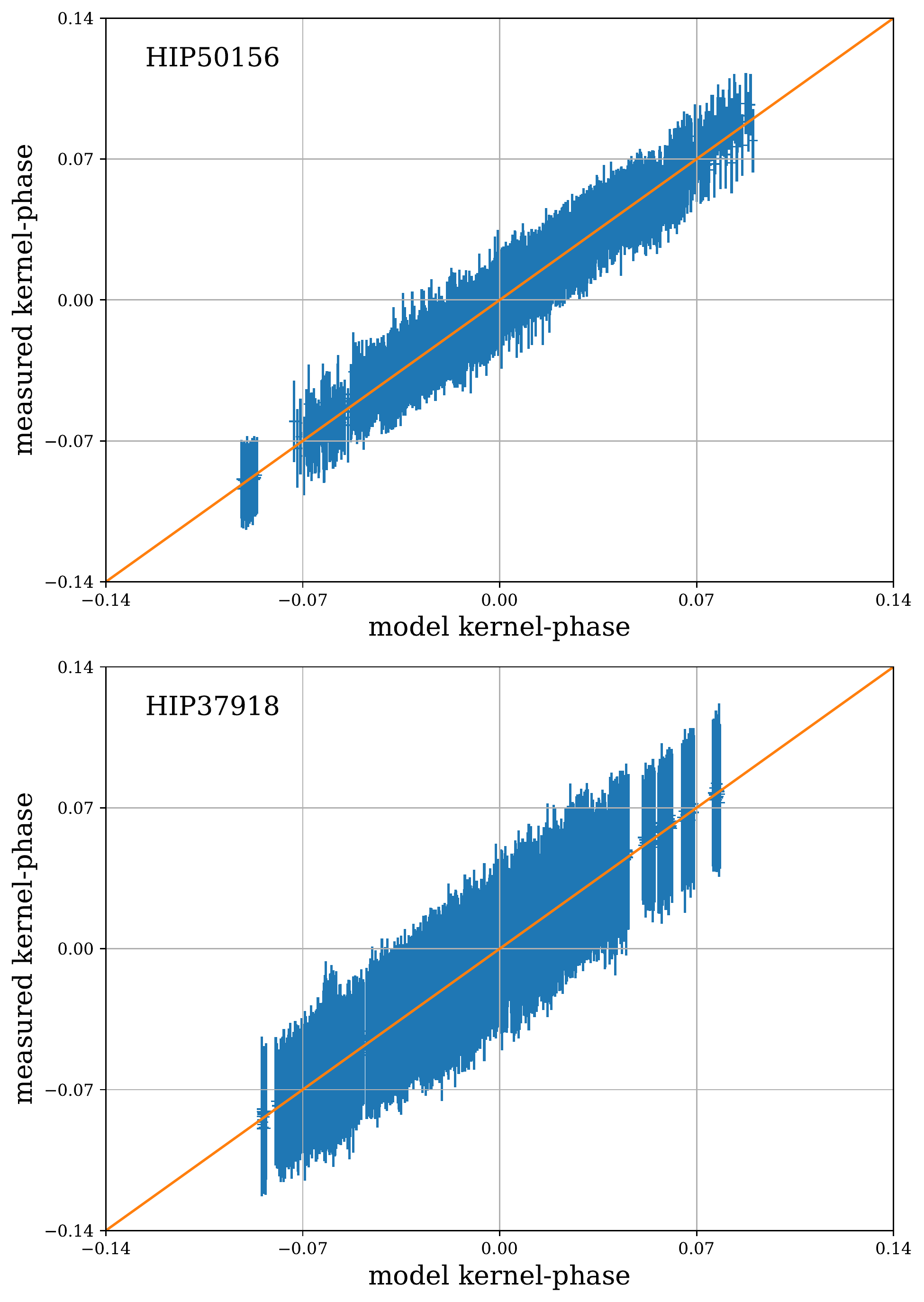}
    \caption{Correlation of the measured kernel phase and the best fit binary model kernel phase for the targets for which we detect a close companion candidate in blue. The presented errorbars are computed based on photon noise (cf. Section~\ref{sec:kernel_phase_uncertainties}) and scaled up by $f_\text{err}$ according to our empirical uncertainties (cf. Section~\ref{sec:empirical_uncertainties}). The orange line indicates the identity which would represent perfect agreement between measured and model kernel phase. Similar to Figure~\ref{fig:hip_47425_kernel_phase_uncertainties} we normalize each kernel phase by the norm of its corresponding row of $\bm{P}'\cdot\bm{K}$ since we are dealing with calibrated kernel phase here.}
    \label{fig:correlation_plots}
\end{figure*}

%\begin{equation}
%    \left.\frac{\partial}{\partial{c}}\chi_\text{bin}^2\right|_{c_\text{fit}} = (-2\cdot\Theta_\text{bin,ref})^T\cdot\bm{\Sigma}_\Theta^{-1}\cdot(\Theta-c_\text{fit}\cdot\Theta_\text{bin,ref}),
%\end{equation}

%\begin{align}
%    \sigma_{c_\text{fit}}^2 &= \frac{(\Theta_\text{bin,ref}^T\cdot\bm{\Sigma}_\Theta^{-1})\cdot\bm{\Sigma}_\Theta\cdot(\Theta_\text{bin,ref}^T\cdot\bm{\Sigma}_\Theta^{-1})^T}{(\Theta_\text{bin,ref}^T\cdot\bm{\Sigma}_\Theta^{-1}\cdot\Theta_\text{bin,ref})^2}\\
%                            &= \frac{\Theta_\text{bin,ref}^T\cdot\bm{\Sigma}_\Theta^{-1}\cdot\Theta_\text{bin,ref}}{(\Theta_\text{bin,ref}^T\cdot\bm{\Sigma}_\Theta^{-1}\cdot\Theta_\text{bin,ref})^2},\\
%\end{align}

%%%%%%%%%%%%%%%%%%%%%%%%%%%%%%%%%%%%%%%%%%%%%%%%%%

% Don't change these lines
\bsp	% typesetting comment
\label{lastpage}
\end{document}